\documentclass[aps,pra,superscriptaddress,twocolumn,eqsecnum]{revtex4-2}
\usepackage{graphicx,color}
\usepackage{amsmath}
\usepackage{amssymb}
\usepackage{amsthm}
\usepackage{multirow}
\usepackage{hyperref}
\usepackage{enumitem}
\usepackage{bm}

\newcommand{\tr}{\mathop{\mathrm{tr}}\nolimits}

\renewcommand{\Re}{\mathop{\mathrm{Re}}\nolimits}

\newcommand{\rank}{\mathop{\mathrm{rank}}\nolimits}
\newcommand{\rmd}{\mathrm{d}}
\newcommand{\rmi}{\mathrm{i}}
\newcommand{\rme}{\mathrm{e}}
\newcommand{\ket}[1]{|{#1}\rangle}
\newcommand{\bra}[1]{\langle{#1}|}

\definecolor{dgreen}{rgb}{0,0.5,0}

\definecolor{dblue}{rgb}{0,0,0.6}

\definecolor{dred}{rgb}{0.784,0,0}

\definecolor{delete}{cmyk}{0.5,0,0,0}

\begin{document}
\title{Eternal Adiabaticity}
\author{Daniel Burgarth} 
\affiliation{Center for Engineered Quantum Systems, Dept.\ of Physics \& Astronomy, Macquarie University, 2109 NSW, Australia}
\author{Paolo Facchi}
\affiliation{Dipartimento di Fisica and MECENAS, Universit\`a di Bari, I-70126 Bari, Italy}
\affiliation{INFN, Sezione di Bari, I-70126 Bari, Italy}
\author{Hiromichi Nakazato}
\affiliation{Department of Physics, Waseda University, Tokyo 169-8555, Japan}
\author{Saverio Pascazio}
\affiliation{Dipartimento di Fisica and MECENAS, Universit\`a di Bari, I-70126 Bari, Italy}
\affiliation{INFN, Sezione di Bari, I-70126 Bari, Italy}
\affiliation{Istituto Nazionale di Ottica (INO-CNR), I-50125 Firenze, Italy}
\author{Kazuya Yuasa}
\affiliation{Department of Physics, Waseda University, Tokyo 169-8555, Japan}
\date{\today}
\begin{abstract}
We iteratively apply a recently formulated adiabatic theorem for the strong-coupling limit in finite-dimensional quantum systems. This allows us to improve approximations to a perturbed dynamics, beyond the standard approximation based on quantum Zeno dynamics and adiabatic elimination.
The effective generators describing the approximate evolutions are endowed with the same block structure as the unperturbed part of the generator, and exhibit adiabatic evolutions.
This iterative adiabatic theorem reveals that  adiabaticity holds \textit{eternally}, that is, the system evolves within each eigenspace of the unperturbed part of the generator, with an error bounded by $O(1/\gamma)$ \textit{uniformly in time}, where $\gamma$ characterizes the strength of the unperturbed part of the generator.
We prove that the iterative adiabatic theorem reproduces Bloch's perturbation theory in the unitary case, 
and is therefore a full generalization to open systems.
We furthermore prove the equivalence of the Schrieffer-Wolff and des Cloiseaux approaches in the unitary case and generalize both to arbitrary open systems, showing that they share the eternal adiabaticity, and
providing explicit error bounds. Finally we discuss the physical structure of the effective adiabatic generators and show that ideal effective generators for open systems do not exist in general.
\end{abstract}
\maketitle

\section{Introduction}
Modeling physical systems is important in physics and science.
Identifying a good effective generator of a system is crucial in the analysis of the physical dynamics of the system.
A separation of time scales is most often a key in doing that.
It allows us to focus on a subset of relevant energy levels of the system.
High-frequency components can be ``adiabatically eliminated,'' and the evolution of the system is well described by an effective generator acting only on the relevant subspace.

Such effective modeling can be justified by an adiabatic theorem~\cite{ref:Messiah,ref:KatoAdiabatic}.
Consider first a closed quantum system with a dynamics dominated by a strong part of its Hamiltonian, and the leakage out of the eigenspaces of the strong Hamiltonian is suppressed due to the separation of time scales.
This ensures that the evolution of the system is well approximated by the adiabatic evolution within the eigenspaces.
In the limit of an infinitely strong separation of time scales, the leakage is completely suppressed and the system is perfectly confined within each eigenspace.
It is known as a version of the quantum Zeno effect~\cite{ref:PaoloAdiabatic,ref:QZS,ref:ControlDecoZeno,ref:PaoloSaverio-QZEreview-JPA}.
The adiabatic evolution within the eigenspaces (quantum Zeno dynamics~\cite{ref:PaoloSaverio-QZEreview-JPA,ref:ArtZeno}) is described by a Hamiltonian projected on the individual eigenspaces (Zeno Hamiltonian).
If on the other hand the separation of time scales is strong but finite, the system can slowly transit between eigenspaces.
An effective Hamiltonian including such processes can be systematically constructed via the technique known as adiabatic elimination~\cite{ref:CohenTannoudjiTextbook-AtomPhoton,Brion}, and refines the approximation by the Zeno Hamiltonian.

In practice, many quantum systems are noisy, and it is important to extend the theory to Lindbladian generators. It is difficult to give the vast literature on this area the deserved attention, and we only provide some exemplary references for such generalizations of the adiabatic theorem~\cite{martin,schmid}, of strong coupling limits~\cite{ref:NoiseInducedZeno,ref:ZanardiDFS-PRL2014,ref:ZanardiDFS-PRA2015,ref:VictorPRX,ref:ZanardiDFS-PRA2017,unity1}, of quantum Zeno dynamics~\cite{unity2,ref:Barankai,ref:Mobus,ref:Becker},
and of adiabatic elimination~\cite{ref:AdiabaticEliminationCiracBlattZollerPhillips,PhysRevA.67.023802,4982633,PhysRevA.85.032111,PhysRevA.101.042102,Sarlette16,Sarlette17,Sarlette18,ref:Sarlette2020}.

All the above theories for effective generators are, however, usually valid for finite time ranges only.
Known error bounds on adiabatic approximations, i.e.~bounds on the distance between the true evolution and an adiabatic evolution within the eigenspaces, grow in time~\cite{ref:KatoAdiabatic,unity1,HamazakiPRL2020,HamazakiPRA2020}, and the adiabaticity of the evolution is not guaranteed by the standard adiabatic theorems in the long term. Accordingly we would need a stronger separation of time scales to realize the adiabatic evolution for a longer time.

In this paper, we show that  adiabaticity actually holds \textit{eternally}.
The system remains within each eigenspace of the strong part of its generator with an error remaining $O(1/\gamma)$ \textit{for arbitrarily long times and arbitrary perturbations}, where $\gamma$ characterizes the strength of the strong Hamiltonian relative to the perturbation.
The reason why the standard adiabatic theorems appear to assure the adiabaticity only for finite times is because the adiabatic generators used in the adiabatic theorems to approximate the true evolutions, e.g.~by Zeno Hamiltonians, are not fine enough.
One can find an adiabatic generator that adapts better to the evolution of the system while provoking no leakage out of the eigenspaces.
It well approximates the true evolution with an error bounded by $O(1/\gamma)$ \textit{uniformly in time}.

Let us summarize the main results of the present work. We consider an evolution $\rme^{t(\gamma B+C)}$ of a finite-dimensional quantum system with a ``strong'' generator $B$ and a ``weak'' generator $C$. These generators can be Hamiltonians or Lindbladians.
In Ref.~\cite{unity1}, we have developed an adiabatic theorem for the strong-coupling limit $\gamma\to+\infty$ for open systems.
Here, we intend to improve the adiabatic approximation by applying the adiabatic theorem iteratively (Sec.~\ref{sec:IteratedAdiabatic}).
This leads us to a good choice of adiabatic generator $\gamma B+D$,  with $D=D(\gamma)$ endowed with the same block structure as $B$, thus provoking no leakage out of the eigenspaces of $B$, and at the same time allowing us to bound the distance
\begin{equation}
\rme^{t(\gamma B+C)}-\rme^{t[\gamma B+D(\gamma)]}= O(1/\gamma)
\label{eqn:EternalAdiabaticityDifference}
\end{equation}
uniformly in time (Sec.~\ref{sec:GoodChoice}).
The $\ell$th block $D_\ell$ of this adiabatic generator $D$ acting on the $\ell$th eigenspace of the strong generator $B$ is given by $D_\ell=P_\ell\Omega_\ell P_\ell$, where $\Omega_\ell$  is a solution of the quadratic operator equation
\begin{equation}
\frac{1}{\gamma}S_\ell \Omega_\ell^2-\left(1+\frac{1}{\gamma}CS_\ell\right)\Omega_\ell
+S_\ell\Omega_\ell N_\ell
+CP_\ell
=0,
\end{equation}
with $\Omega_\ell=\Omega_\ell P_\ell$, 
and $P_\ell$ and $N_\ell$ are the spectral projection and the nilpotent of the $\ell$th eigenspace of $B$, respectively, while $S_\ell$ is the reduced resolvent of $B$ at its $\ell$th eigenvalue~\cite{ref:KatoBook} (their details are provided in the following section).
We realize that $U_\ell=P_\ell-S_\ell\Omega_\ell/\gamma$ satisfies another quadratic equation
\begin{equation}
U_\ell
-S_\ell U_\ell N_\ell
+\frac{1}{\gamma}S_\ell(CU_\ell-U_\ell CU_\ell)
-P_\ell
=0,
\end{equation}
with $U_\ell P_\ell=U_\ell$ (Appendix~\ref{app:GeneralizedBlochEq}), and in the absence of the nilpotent $N_\ell$ in the unitary case this equation is nothing but the well-known Bloch equation~\cite{Bloch,Lindgren}.
The iterated adiabatic theorem thus reproduces Bloch's perturbation theory developed for closed systems~\cite{Bloch,Cloizeaux,Lindgren,Klein,Killingbeck}, and it is here generalized to open systems.
Although we also provide perturbative expansions (Sec.~\ref{Perturbative}), our key focus is the adiabatic generator $D=\sum_\ell D_\ell$, whose components $D_\ell(\gamma)$ are a resummation of a full-order perturbative series.
We show the nonperturbative solvability of the Bloch equation and the region where the relevant solution exists and is unique (Appendix~\ref{bounds}) using the Newton-Kantorovich theorem~\cite{Ortega}.
This allows us to explicitly bound the eternal adiabaticity~(\ref{eqn:EternalAdiabaticityDifference}) (Sec.~\ref{boundingall} and Appendix~\ref{app:EternalBounds}).

Next we turn our attention to the structure of the effective generator. Behind  eternal adiabaticity, we have similarity
\begin{equation}
\gamma B+C=U(\gamma B+D)U^{-1}
\end{equation}
between the adiabatic generator $\gamma B+D$ and the original generator $\gamma B+C$, with $U=\sum_\ell U_\ell=1+ O(1/\gamma)$ (see Sec.~\ref{similarity}). It is known, however, that even in the unitary case there is a lot of gauge freedom in the choice of good adiabatic generators.
This fact encourages us to take an axiomatic approach to define an \emph{ideal} effective adiabatic generator, as initiated for the unitary case in Ref.~\cite{Soliverez}:
\begin{enumerate}
\item
An effective adiabatic generator $C_\text{eff}$ should be endowed with the same block structure as $B$, i.e., $[C_\text{eff},P_\ell]=0$, provoking no leakage out of the eigenspaces of $B$.

\item
The effective adiabatic generator $\gamma B+C_\text{eff}$ should be similar to the original generator $\gamma B+C$, sharing the same spectrum.

\item
The similarity transformation $U$ should be small, i.e., close to the identity, $U=1+ O(1/\gamma)$.

\item
The effective adiabatic generator $\gamma B+C_\text{eff}$ should be physical, i.e., Hermiticity-preserving (HP), trace-preserving (TP), and conditionally completely positive (CP) (with a positive-semidefinite Kossakowski matrix)~\cite{ref:VectorizationHavel}, generating a completely positive evolution~\cite{ref:DynamicalMap-Alicki,ref:GKLS-DariuszSaverio}.
\end{enumerate}

While the first three axioms suffice to show eternal adiabaticity, the fourth is desirable to get a direct physical interpretation of the generator. It is known in the literature that, due to an asymmetry in the construction, the adiabatic generator $D$ from Bloch's perturbation theory is not skew-Hermitian in general even in the unitary case with skew-Hermitian $B$ and $C$~\cite{Bloch,Cloizeaux,Lindgren,Klein,Shavitt,Killingbeck,Bravyi}.
In the unitary case, on the other hand, des Cloizeaux showed that one can turn the non-skew-Hermitian $\gamma B+D$ into a skew-Hermitian $\gamma B+K$ by an additional similarity transformation keeping the block structure~\cite{Cloizeaux,Klein}. This is an example of an ideal effective generator.

A skew-Hermitian effective generator on a particular eigenspace (without caring about the block structure of the other eigenspaces) can also be obtained from the original $\gamma B+C$ via the Schrieffer-Wolff transformation in the unitary case~\cite{Schrieffer,Shavitt,Bravyi}. The connection between Schrieffer-Wolff's, adiabatic elimination, and des Cloizeaux's perturbative approaches has been noted before~\cite{Sanz}, and another higher-order adiabatic elimination based on a Lippmann-Schwinger-type equation was derived~\cite{Englert2,Englert1}.

The generalization of Schrieffer-Wolff transformations to open systems was investigated in Ref.~\cite{Kessler}, where the author focused on the stationary subspace, i.e.~the eigenspace of $B$ belonging to the eigenvalue $0$, and assumed that the generator $B$ is diagonalizable, with no nilpotent. Physicality was analyzed up to the third order for some specific settings.

Here, based on our generalization of Bloch's equation, we provide a nonperturbative generalization of the Schrieffer-Wolff and des Clouiseaux approaches to the open-system  case (Secs.~\ref{sec:ConjugateBloch} and~\ref{sec:SW}). We construct a very natural and symmetric similarity transformation from the solutions of Bloch's equation which fulfills the first three axioms of an ideal effective generator and reduces to the des Clouiseaux approach in the unitary case. Our formalism can be applied to general generators, which are not necessarily diagonalizable and can admit nilpotents, and deals with all the eigenspaces, including the nonstationary ones, respecting the block structure. We prove that the adiabatic generators are both HP and TP for general open systems (Sec.~\ref{sec:Physicality}).

After providing a general framework, we will look at a few examples in Sec.~\ref{sec:Examples}: a dissipative $\Lambda$ system, for which an analytical expression for the nonperturbative (full-order) adiabatic generator is available (Sec.~\ref{sec:DissipativeLambda}), and a system admitting a nilpotent in the strong part $B$ (Sec.~\ref{sec:ExampleNilpotent}). We find that our effective generator is not always completely positive (that is, the fourth axiom is not always fulfilled).

Could there be another approach (choice of gauge) which fulfills all axioms? Surprisingly we show that this is generally impossible by providing a counterexample (Sec.~\ref{sec:Impossibility}) in which axioms one and two imply breaking axiom four.
If one wishes to require that an effective generator for an open system should have the complete physical structure (HP, TP, and CP), as a trade-off axioms one and/or two in the above list should be abandoned. There are attempts to develop a general perturbation theory along those lines~\cite{Sarlette16,Sarlette17,Sarlette18,ref:Sarlette2020}.

We will conclude the paper in Sec.~\ref{sec:Conclusions} and provide some details in Appendices~\ref{app:KeyFormula}--\ref{app:EternalBounds}\@.

Here, we take the view that the eternal adiabaticity is the most striking feature, as it highlights a certain robustness of quantum evolutions against perturbations. This aspect is further elaborated in Ref.~\cite{shorty}, where we explore connections to KAM stability. 

\section{Iterated Adiabatic Theorem}
\label{sec:IteratedAdiabatic}
We iteratively apply the adiabatic theorem developed in Ref.~\cite{unity1}, to improve the adiabatic approximation.
The goal is to find a good approximation of $\rme^{t(\gamma B+C)}$ by $\rme^{t(\gamma B+D)}$ with an operator $D$ endowed with the same block structure as $B$, causing no leakage from each eigenspace of $B$.
We will show that there exists such a generator $D$ that ensures that the error of $\rme^{t(\gamma B+D)}$ to $\rme^{t(\gamma B+C)}$ remains $O(1/\gamma)$ for arbitrarily long times $t$.
Essentially, one can think of this approach as a type of perturbation theory within the exponential function. 

Although we ultimately have physical operators (Hamiltonians and Lindbladians) in mind, most of the results of this paper are valid for \emph{arbitrary} square matrices $B$ and $C$, \emph{without} requiring \emph{any structural assumptions} on them.

Let
\begin{equation}
B=\sum_\ell(b_\ell P_\ell+N_\ell)
\end{equation}
be the \textit{canonical form} or the \textit{spectral representation} of $B$ (recall the \textit{Jordan normal form})~\cite{ref:KatoBook}.
Here, 
$\{b_\ell\}$ is the \textit{spectrum} of $B$, which is the set of distinct eigenvalues of $B$ (labeled such that $b_k\neq b_\ell$ for $k\neq \ell$), $\{P_\ell\}$ are the corresponding eigenprojections, called the \textit{spectral projections} of $B$, satisfying
\begin{equation}
\label{eq:PVM}
P_kP_\ell=\delta_{k\ell}P_k,\qquad\sum_\ell P_\ell=1,
\end{equation}
for all $k$ and $\ell$, and $\{N_\ell\}$ are the corresponding \textit{nilpotents} of $B$, satisfying
\begin{equation}
\label{eq:nilpot}
P_k N_\ell = N_\ell P_k = \delta_{k\ell} N_k, \qquad N_\ell^{n_\ell} = 0,
\end{equation}
for all $k$ and $\ell$, and for some integers $1\leq n_\ell\leq \rank P_\ell$. Notice that the spectral projections, which determine the partition of the space through the \textit{resolution of identity}~(\ref{eq:PVM}), are not Hermitian in general, $P_\ell\neq P_\ell^\dag$. 
We set
\begin{equation}
B_\ell=BP_\ell=b_\ell P_\ell+N_\ell.
\end{equation}

First, we focus on a particular eigenspace of $B$ belonging to eigenvalue $b_\ell$, and find a suitable $D_\ell$ that describes the adiabatic evolution of the system in the eigenspace for large $\gamma$.
The following iteration works for any choice of $D_\ell$ satisfying
\begin{equation}
D_\ell = P_\ell D_\ell P_\ell,
\label{eq:Dellcond}
\end{equation}
and hence having the same block structure as $B$.
However later we will find out that there are particularly good choices of $D_\ell$.

We wish to estimate the difference between $\rme^{t(\gamma B+C)}	P_\ell$ and $\rme^{t(\gamma B+D_\ell)} P_\ell$.
It can be estimated by writing it as an integral:
\begin{align}
&\bigl(
\rme^{t(\gamma B+C)}
-\rme^{t(\gamma B+D_\ell )}
\bigr)
P_\ell
\nonumber\\
&\quad
 =-\int_0^t\rmd s\,\frac{\partial }{\partial s}(
\rme^{(t-s)(\gamma B+C)}
\rme^{s(\gamma B+D_\ell)}
)
P_\ell
\nonumber
\displaybreak[0]\\
&\quad
=\int_0^t\rmd s\,
\rme^{(t-s)(\gamma B+C)}
(C-D_\ell )
P_\ell
\rme^{s(\gamma B+D_\ell )}.
\label{eqn:difference0}
\end{align}
The key quantity from Ref.~\cite{unity1} is the reduced resolvent $S_\ell$, defined by
\begin{equation}
S_\ell=\sum_{k\neq\ell}(b_k-b_\ell+N_k)^{-1}P_k
\label{eqn:Resolvent}
\end{equation}
(see Refs.~\cite{ref:KatoAdiabatic,ref:KatoBook} for the unitary case).
This satisfies
\begin{gather}
P_\ell S_\ell=S_\ell P_\ell=0,\\
(B-b_\ell )S_\ell
=S_\ell(B-b_\ell )
=1-P_\ell.
\label{eqn:BSP}
\end{gather}
In addition, the key formula for the adiabatic theorem is given by
\begin{align}
&\int_0^t\rmd s\,
\rme^{(t-s)(\gamma B+C)}
A P_\ell
\rme^{s(\gamma B+D_\ell)}
\nonumber
\\
&\quad
=
\int_0^t\rmd s\,
\rme^{(t-s)(\gamma B+C)}
P_\ell
A
P_\ell
\rme^{s(\gamma B+D_\ell )}
\nonumber\\
&\qquad{}
+\frac{1}{\gamma}
\rme^{t(\gamma B+C)}
S_\ell
A
P_\ell
-\frac{1}{\gamma}
S_\ell
A
P_\ell
\rme^{t(\gamma B+D_\ell )}
\nonumber\\
&\qquad{}
-\frac{1}{\gamma}
\int_0^t\rmd s\,
\rme^{(t-s)(\gamma B+C)}
\mathcal{K}_\ell(A)
P_\ell
\rme^{s(\gamma B+D_\ell )},
\label{eqn:KeyRelation}
\end{align}
where
\begin{equation}
\mathcal{K}_\ell(A)=CS_\ell A-S_\ell AD_\ell-\gamma S_\ell AN_\ell,
\label{eq:Kelldef}
\end{equation}
for an arbitrary operator $A$.
See Appendix~\ref{app:KeyFormula} for the derivation of this key formula.
Then, the difference~(\ref{eqn:difference0}) can be immediately estimated by applying the key formula~(\ref{eqn:KeyRelation}) for $A=C-D_\ell\equiv A_\ell^{(0)}$.
In particular, if $D_\ell$ is chosen to be $D_\ell=P_\ell CP_\ell$, then $P_\ell A_\ell^{(0)}P_\ell=P_\ell(C-D_\ell)P_\ell=0$ and the first integral in the right-hand side identically vanishes. Moreover, if there is no nilpotent $N_\ell=0$ in the relevant eigenspace, then $\mathcal{K}_\ell$ is independent of $\gamma$, and we get 
\begin{align}
&
\bigl(
\rme^{t(\gamma B+C)}
-\rme^{t(\gamma B+P_\ell CP_\ell)}
\bigr)
P_\ell
\nonumber\\
&\quad
=
\frac{1}{\gamma}
\bigl(
\rme^{t(\gamma B+C)}
S_\ell
CP_\ell
-
S_\ell
CP_\ell
\rme^{t(\gamma B+P_\ell CP_\ell)}
\bigr)
\nonumber
\\
&\qquad
{}-\frac{1}{\gamma}
\int_0^t\rmd s\,
\rme^{(t-s)(\gamma B+C)}
[
C
,
S_\ell
C
P_\ell
]
P_\ell
\rme^{s(\gamma B+P_\ell CP_\ell)}.
\label{eqn:Adiabatic1st}
\end{align}
This provides an adiabatic theorem~\cite{unity1}: when $B$ is Lindbladian or Hamiltonian, so that the semigroup it generates is uniformly bounded in time, then the evolution is confined within the eigenspace specified by its spectral projection $P_\ell$, with an error $O(1/\gamma)$ for any finite $t$.
The adiabatic evolution within the eigenspace is described by the generator $D_\ell=P_\ell CP_\ell$.
However, the error would accumulate by the last integral as time $t$ goes on, and the above adiabatic theorem~(\ref{eqn:Adiabatic1st}) does not ensure the adiabaticity of the evolution for long times of $O(\gamma)$. See e.g.\ Fig.~\ref{fig:DissipativeLambdaEvolution} below.

Still, with a careful choice of the generator $D_\ell$, one can ensure the adiabaticity to hold \textit{eternally}, for arbitrarily long times.
We are going to show this by iteratively refining the generator $D_\ell$, and so pushing the validity of the adiabatic approximation to times of higher and higher order of $\gamma$.

To improve the approximation, we iteratively apply the key formula~(\ref{eqn:KeyRelation}), to the last integral in its right-hand side.
After $n$ iterations, we get
\begin{align}
&
\bigl(
\rme^{t(\gamma B+C)}
-\rme^{t(\gamma B+D_\ell )}
\bigr)
P_\ell
\nonumber\\
&%
=
\int_0^t\rmd s\,
\rme^{(t-s)(\gamma B+C)}
\,\biggl(
\sum_{j=0}^n
\frac{(-1)^j}{\gamma^j}
P_\ell
A_\ell^{(j)}
P_\ell
\biggr)\,
\rme^{s(\gamma B+D_\ell )}
\nonumber
\displaybreak[0]
\\
&\quad
{}+\frac{1}{\gamma}
\rme^{t(\gamma B+C)}
\,\biggl(
\sum_{j=0}^{n-1}
\frac{(-1)^j}{\gamma^j}
S_\ell
A_\ell^{(j)}
P_\ell
\biggr)
\nonumber\
\\
&\qquad\qquad\ \ \,
{}
-\frac{1}{\gamma}
\,\biggl(
\sum_{j=0}^{n-1}
\frac{(-1)^j}{\gamma^j}
S_\ell
A_\ell^{(j)}
P_\ell
\biggr)\,
\rme^{t(\gamma B+D_\ell )}
\nonumber
\displaybreak[0]
\\
&\quad
{}+\frac{(-1)^n}{\gamma^n}
\int_0^t\rmd s\,
\rme^{(t-s)(\gamma B+C)}
A^{(n)}_\ell P_\ell
\rme^{s(\gamma B+D_\ell)},
\label{key}
\end{align}
where
\begin{equation}
A^{(0)}_\ell=C-D_\ell, \ \ %
A^{(n)}_\ell=\mathcal{K}_\ell(A^{(n-1)}_\ell) = \mathcal{K}_\ell^n(A^{(0)}_\ell).
\label{recursion}
\end{equation}
As proved in Appendix~\ref{app:Bounding2.16}\@, if
\begin{equation}
\gamma>\max\{1,[\|S_\ell\|(\|C\|+\|D_\ell\|+\|N_\ell\|)]^{n_\ell}\},
\end{equation}
then the last contribution in~(\ref{key}) decays out exponentially as $n\to+\infty$
and the series
\begin{equation}
G_\ell=\sum_{j=0}^\infty\frac{(-1)^j}{\gamma^j} A_\ell^{(j)} =\sum_{j=0}^\infty\frac{(-1)^j}{\gamma^j}\mathcal{K}_\ell^j( C-D_\ell)
\label{noequationnumber}
\end{equation}
converges. Here and in the following, we will consider only unitary invariant norms, with the spectral and trace norms the most useful.
Thus, in the limit $n\rightarrow+\infty$ one gets
\begin{align}
&\bigl(
\rme^{t(\gamma B+C)}-\rme^{t(\gamma B+D_\ell)}
\bigr)
P_\ell
\nonumber\\
&\quad
=\int_0^t\rmd s\,\rme^{(t-s)(\gamma B+C)}P_\ell G_\ell P_\ell \rme^{s(\gamma B+D_\ell)}
\nonumber
\\
&\qquad{}
+\frac{1}{\gamma}\bigl(\rme^{t(\gamma B+C)}S_\ell G_\ell P_\ell-S_\ell G_\ell P_\ell \rme^{t(\gamma B+D_\ell)}\bigr).
\label{eqn:IteratedAdiabaticLimit0}
\end{align}
This equation holds for any choice of $D_\ell$  with the same block structure as $B$ as in~\eqref{eq:Dellcond}, and for any sufficiently large $\gamma$. 
We now seek a $D_\ell $ such that 
\begin{equation}
P_\ell G_\ell P_\ell=0,
\label{eq:PlGl}
\end{equation} 
so that the last integral, which would grow in time and make the error bound larger and larger, vanishes, giving
\begin{align}
&\bigl(
\rme^{t(\gamma B+C)}-\rme^{t(\gamma B+D_\ell)}
\bigr)
P_\ell
\nonumber\\
&\qquad{}
=\frac{1}{\gamma}\bigl(\rme^{t(\gamma B+C)}S_\ell G_\ell P_\ell-S_\ell G_\ell P_\ell \rme^{t(\gamma B+D_\ell)}\bigr).
\label{eqn:IteratedAdiabaticLimit}
\end{align}
Such a $D_\ell$ actually exists, as proved in the next section.

\section{Adiabatic Bloch Equation}
\label{sec:GoodChoice}
The adiabatic generator $D_\ell$ fulfilling the condition~\eqref{eq:PlGl} and thus giving~(\ref{eqn:IteratedAdiabaticLimit}) is  given by
\begin{equation}
D_\ell =P_\ell\Omega_\ell = P_\ell\Omega_\ell P_\ell,
\label{eqn:Omega2D}
\end{equation}
where $\Omega_\ell$ is a solution of the quadratic equation
\begin{equation}
\frac{1}{\gamma}S_\ell \Omega_\ell^2-\left(1+\frac{1}{\gamma}CS_\ell\right)\Omega_\ell
+S_\ell\Omega_\ell N_\ell
+CP_\ell
=0,
\label{eqn:AdiabaticBloch}
\end{equation}
with 
\begin{equation}
\Omega_\ell(1- P_\ell) =0.
\label{eqn:AB2}
\end{equation}
Because this equation is derived from the iterated adiabatic theorem, and because it generalizes the well-known Bloch wave operator equation~\cite{Bloch,Lindgren} as shown in Appendix~\ref{app:GeneralizedBlochEq}, we call the quadratic equation~(\ref{eqn:AdiabaticBloch}) with~(\ref{eqn:AB2}) for $\Omega_\ell$ the \emph{adiabatic Bloch equation}. 

With such a particular choice of $D_\ell$, we have that  $S_\ell G_\ell P_\ell=S_\ell\Omega_\ell=S_\ell\Omega_\ell P_\ell$, and  Eq.~(\ref{eqn:IteratedAdiabaticLimit}) reduces to
\begin{multline}
\bigl(
\rme^{t(\gamma B+C)}-\rme^{t(\gamma B+D_\ell )}
\bigr)
P_\ell
\\
=
\frac{1}{\gamma}\bigl(
\rme^{t(\gamma B+C)}S_\ell\Omega_\ell P_\ell -S_\ell\Omega_\ell P_\ell \rme^{t(\gamma B+D_\ell)}
\bigr).
\label{eqn:EternalAdiabaticLimit}
\end{multline}
This is valid for arbitrary operators $B$ and $C$, not necessarily Hamiltonians or Lindbladians.

\subsection{Derivation of the Adiabatic Bloch Equation}
Let us start by looking at the condition~\eqref{eq:PlGl}.
For large enough $\gamma$, the series~(\ref{noequationnumber}) converges, the inverse $(1+\gamma^{-1} \mathcal{K}_\ell)^{-1}$ exists, and we get
\begin{equation}
G_\ell=(1+\gamma^{-1} \mathcal{K}_\ell)^{-1}(C- D_\ell).
\label{eqn:G0}
\end{equation}
By the block structure of $D_\ell$ in~\eqref{eq:Dellcond} and by using $S_\ell P_\ell=0$, one gets $\mathcal{K}_\ell(D_\ell) = 0$, whence
\begin{equation}
	G_\ell=(1+\gamma^{-1} \mathcal{K}_\ell)^{-1}(C) - D_\ell.
	 \label{eqn:G}
\end{equation}
Since $\mathcal{K}_\ell (A) P_\ell = \mathcal{K}_\ell (A P_\ell)$ and $D_\ell=P_\ell D_\ell P_\ell$, the condition $P_\ell G_\ell P_\ell=0$ is equivalent to 
\begin{equation}
P_\ell G_\ell P_\ell
=P_\ell(1+\gamma^{-1} \mathcal{K}_\ell)^{-1}(C P_\ell) - D_\ell
=0,
\end{equation}
which in turn implies
\begin{equation}
(1+\gamma^{-1} \mathcal{K}_\ell)^{-1}(C P_\ell) - D_\ell = R_\ell,
\end{equation}
with $R_\ell = (1- P_\ell) R_\ell P_\ell$.
Then, by setting $\Omega_\ell=D_\ell+R_\ell = \Omega_\ell P_\ell$, it reads
\begin{equation}
(1+\gamma^{-1} \mathcal{K}_\ell)^{-1}(C P_\ell) = \Omega_\ell.
\label{eq:3.8}
\end{equation}
By inverting,
\begin{equation}
C P_\ell  =   \Omega_\ell  +  \frac{1}{\gamma}\mathcal{K}_\ell (\Omega_\ell),
\end{equation}
that is, by the definition~\eqref{eq:Kelldef} of~$\mathcal{K}_\ell$,
\begin{equation}
CP_\ell
=\Omega_\ell  
+\frac{1}{\gamma}CS_\ell\Omega_\ell 
-\frac{1}{\gamma}S_\ell\Omega_\ell D_\ell
-S_\ell\Omega_\ell N_\ell.
\end{equation}
Since $\Omega_\ell R_\ell=0$, we can write $\Omega_\ell D_\ell=\Omega_\ell^2$.
Therefore, we get the quadratic equation~(\ref{eqn:AdiabaticBloch}) for $\Omega_\ell$ with~(\ref{eqn:AB2}).

It follows from the Newton-Kantorovich theorem~\cite{Ortega} that for large enough $\gamma$ the adiabatic Bloch equation~(\ref{eqn:AdiabaticBloch}) with~(\ref{eqn:AB2}) has a unique solution within a certain range.
See Appendix~\ref{bounds}\@. 
From such a solution $\Omega_\ell$, we obtain the wanted $D_\ell$ by~(\ref{eqn:Omega2D}).

\subsection{Simplifying $G_\ell$}
The solution of the adiabatic Bloch equation~(\ref{eqn:AdiabaticBloch}) with~(\ref{eqn:AB2}) allows us to simplify the expression for $G_\ell$.
To this end, let us look at the components of $G_\ell$ other than $P_\ell G_\ell P_\ell$, which vanishes by~\eqref{eq:PlGl}.
From~(\ref{eqn:G}) and~(\ref{eq:3.8}), we get
\begin{align}
(1-P_\ell)G_\ell P_\ell 
&=(1-P_\ell)(1+\gamma^{-1} \mathcal{K}_\ell)^{-1}(C P_\ell) 
\nonumber\\
&=(1-P_\ell)\Omega_\ell P_\ell,
\end{align}
where we have used $(1-P_\ell)D_\ell=0$ and $\mathcal{K}_\ell (A) P_\ell = \mathcal{K}_\ell(A P_\ell)$. 
Therefore, we get $S_\ell G_\ell P_\ell=S_\ell\Omega_\ell P_\ell$ and Eq.~(\ref{eqn:IteratedAdiabaticLimit}) reduces to~(\ref{eqn:EternalAdiabaticLimit}).

In summary, our key equation is the adiabatic Bloch equation~(\ref{eqn:AdiabaticBloch}) with~(\ref{eqn:AB2}).
It admits a unique solution $\Omega_\ell$ within a certain range for large enough $\gamma$ (Appendix~\ref{bounds}).
A good choice of $D_\ell$ describing the adiabatic evolution within the relevant eigenspace is given by~(\ref{eqn:Omega2D}), with which the difference between the adiabatic evolution and the true evolution is estimated as~(\ref{eqn:EternalAdiabaticLimit}).

\section{Perturbative Solution of the Adiabatic Bloch Equation}
\label{Perturbative}
Let us look for a perturbative solution of the adiabatic Bloch equation~(\ref{eqn:AdiabaticBloch}) with~(\ref{eqn:AB2}) in the form
\begin{equation}
	 \Omega_\ell =\Omega_\ell^{(0)}+\frac{1}\gamma\Omega_\ell^{(1)}+\frac{1}{\gamma^2}\Omega_\ell^{(2)}+\cdots=\sum_{j=0}^\infty\frac{1}{\gamma^j}\Omega_\ell^{(j)}.
\end{equation}
Substituting it into the adiabatic Bloch equation~(\ref{eqn:AdiabaticBloch}) and comparing order by order, we obtain  
\begin{gather}
\Omega_\ell^{(0)}-S_\ell \Omega_\ell^{(0)}N_\ell=CP_\ell,
\label{eqn:Iterative0}
\displaybreak[0]
\\
\Omega_\ell^{(j)}
-S_\ell\Omega_\ell^{(j)}N_\ell
=
-CS_\ell\Omega_\ell^{(j-1)}
+S_\ell\sum_{i=0}^{j-1}\Omega_\ell^{(j-i-1)}\Omega_\ell^{(i)}.
\label{iteration}
\end{gather}
By solving this iterative equation, we get that $\Omega_\ell^{(j)} = \Omega_\ell^{(j)} P_\ell$ and  the perturbative expressions for $D_\ell^{(j)}=P_\ell\Omega_\ell^{(j)}$ read
\begin{align}
D_\ell^{(0)}={}&P_\ell CP_\ell,
\label{eqn:D0}
\\
D_\ell^{(1)}
={}&
{-P_\ell}CS_\ell\langle C\rangle P_\ell,
\label{eqn:D1}
\\
D_\ell^{(2)}
={}&
P_\ell
CS_\ell\langle CS_\ell\langle C\rangle\rangle
P_\ell
-
P_\ell
CS_\ell^2\langle\langle C\rangle P_\ell C\rangle
P_\ell,
\label{eqn:D2}
\displaybreak[0]
\\
D_\ell^{(3)}
={}&
{-P_\ell}CS_\ell\langle
CS_\ell\langle
CS_\ell\langle C\rangle
\rangle
\rangle
P_\ell
\nonumber
\displaybreak[0]
\\
&{}
+
P_\ell CS_\ell\langle
CS_\ell^2\langle\langle C\rangle P_\ell C\rangle
\rangle
P_\ell
\nonumber
\displaybreak[0]
\\
&{}+
P_\ell CS_\ell^2\langle\langle C\rangle
P_\ell CS_\ell\langle C\rangle
\rangle
P_\ell
\nonumber\\
&{}
+
P_\ell CS_\ell^2\langle\langle
CS_\ell\langle C\rangle
\rangle
P_\ell
C
\rangle
P_\ell
\nonumber\\
&{}
-
P_\ell CS_\ell^3\langle\langle\langle C\rangle P_\ell C\rangle P_\ell
C\rangle
P_\ell,
\label{eqn:D3}
\end{align}
where we set
\begin{equation}
\langle A\rangle
=\sum_{n=0}^{n_\ell-1}S_\ell^nAN_\ell^n,
\label{eqn:Sandwitch}
\end{equation}
for an arbitrary operator $A$.
If there is no nilpotent $N_\ell$ (i.e.~$n_\ell=1$) in the relevant eigenspace, we simply have $\langle A\rangle=A$, and these expressions reproduce
the perturbative series obtained in Refs.~\cite{Bloch,Cloizeaux}, but are here generalized to nonunitary evolution.

Notice that the zeroth-order term $D_\ell^{(0)}$ in~(\ref{eqn:D0}) is nothing but the ``Zeno generator''~\cite{ref:QZS,ref:PaoloSaverio-QZEreview-JPA,unity1,ref:ArtZeno}, while the first-order term $D_\ell^{(1)}$ yields the ``adiabatic elimination''~\cite{ref:CohenTannoudjiTextbook-AtomPhoton,Brion,ref:AdiabaticEliminationCiracBlattZollerPhillips,PhysRevA.101.042102}.
The higher-order terms refine the approximation beyond the adiabatic elimination.

\section{Similarity of the  Generators}
\label{similarity}
Let us gather the adiabatic generators $D_\ell=P_\ell\Omega_\ell P_\ell$ and define
\begin{equation}
D=\sum_\ell D_\ell.
\end{equation}
The total generator $\gamma B+D$ describing the adiabatic evolution of the system within the eigenspaces is similar to the original generator $\gamma B+C$.
That is, the intertwining relations 
\begin{equation}
(\gamma B+C)U_\ell=U_\ell(\gamma B+D_\ell)
\label{eqn:UD=HU}
\end{equation}
hold for all the operators
\begin{equation}
U_\ell
=P_\ell-\frac{1}{\gamma}S_\ell\Omega_\ell P_\ell,
\label{eqn:Uell}
\end{equation}
and this implies the similarity relation
\begin{equation}
\gamma B+D=U^{-1}(\gamma B+C)U,
\label{eqn:Similarity}
\end{equation}
for sufficiently large $\gamma$,
where
\begin{equation}
U
=\sum_\ell U_\ell
=
1-\frac{1}{\gamma}\sum_\ell S_\ell\Omega_\ell P_\ell.
\label{U}
\end{equation}
Let us prove these facts in this section.
We will use the properties
\begin{equation}
U_\ell P_\ell=U_\ell,\qquad
P_\ell U_\ell=P_\ell.
\label{eqn:UPPU}
\end{equation}

\subsection{Intertwining Relations}
By using the definition of $U_\ell$ in~(\ref{eqn:Uell}), we have
\begin{align}
&(\gamma B+C-\gamma b_\ell )U_\ell
\nonumber\displaybreak[0]\\
&\quad
=
\gamma N_\ell
+CP_\ell
-\frac{1}{\gamma}
(\gamma B+C-\gamma b_\ell )
S_\ell\Omega_\ell.
\intertext{Recalling that $(B-b_\ell )S_\ell=1-P_\ell$ in~(\ref{eqn:BSP}),}
&\quad
=
\gamma N_\ell
+CP_\ell
-
(1-P_\ell)
\Omega_\ell
-\frac{1}{\gamma}
C
S_\ell\Omega_\ell.
\intertext{Using the adiabatic Bloch equation~(\ref{eqn:AdiabaticBloch}),}
&\quad
=
\gamma N_\ell
+P_\ell
\Omega_\ell
-\frac{1}{\gamma}
S_\ell\Omega_\ell^2
-
S_\ell\Omega_\ell N_\ell
\nonumber\displaybreak[0]\\
&\quad
=
\left(
P_\ell
-\frac{1}{\gamma}
S_\ell\Omega_\ell\right)
(
P_\ell\Omega_\ell
+
\gamma N_\ell
)
\nonumber\displaybreak[0]\\
&\quad
=
U_\ell(D_\ell+\gamma N_\ell).
\vphantom{\frac{1}{\gamma}}
\end{align}
Finally, since $U_\ell=U_\ell P_\ell$ and $P_\ell B=P_\ell(b_\ell +N_\ell)$, this gives~(\ref{eqn:UD=HU}).

\subsection{Similarity of the Generators}
Summing the intertwining relations in~(\ref{eqn:UD=HU}) over $\ell$ and noting $U_\ell=U_\ell P_\ell$,
\begin{align}
(\gamma B+C)U
&=\sum_\ell(\gamma B+C)U_\ell	
\nonumber
\\
&=\sum_\ell U_\ell(\gamma B+D_\ell)\nonumber
\\
&=
\sum_\ell U_\ell(\gamma B+D)
\nonumber
\displaybreak[0]
\\
&=U(
\gamma B+D
).
\end{align}
This proves the similarity relation~(\ref{eqn:Similarity}).

The operator $U_\ell$ reduces to Bloch's wave operator~\cite{Bloch,Lindgren} in the unitary case, as shown in Appendix~\ref{app:GeneralizedBlochEq}\@.
Here, it is generalized to open systems, where $B$ can have nilpotents.
One can prove that $U_\ell$ is a solution of the equation
\begin{equation}
U_\ell
-S_\ell U_\ell N_\ell
+\frac{1}{\gamma}S_\ell(CU_\ell-U_\ell CU_\ell)
-P_\ell
=0,
\label{eqn:Bloch}
\end{equation}
with
\begin{equation}
U_\ell (1- P_\ell)= 0 .
\label{eqn:UP}
\end{equation}
See Appendix~\ref{app:GeneralizedBlochEq} for the derivation.
Compared with the original Bloch equation~\cite{Bloch}, the equation~(\ref{eqn:Bloch}) contains an additional term that takes care of the nilpotent $N_\ell$.

We are mainly interested in the evolutions of physical systems, but the similarity and the generalized Bloch equation discussed here are valid for arbitrary operators $B$ and $C$, not necessarily Hamiltonians or Lindbladians.

\section{Eternal Adiabaticity}
\label{boundingall} 
The similarity~(\ref{eqn:Similarity}) proved in the previous section allows us to reproduce the relation~(\ref{eqn:EternalAdiabaticLimit})  immediately. 
Indeed, the similarity~(\ref{eqn:Similarity}) of the generators implies the similarity of the evolutions,
\begin{equation}
\rme^{t(\gamma B+C)}U=U\rme^{t(\gamma B+D)}.
\end{equation}
By inserting the definition of $U$ in~(\ref{U}), we get
\begin{multline}
\rme^{t
(\gamma B+C)
}
-\rme^{t(\gamma B+D)}
\\
=
\frac{1}{\gamma}
\sum_\ell
\bigl(
\rme^{t(\gamma B+C)}
S_\ell\Omega_\ell P_\ell
-
S_\ell\Omega_\ell P_\ell
\rme^{t(\gamma B+D)}
\bigr).
\label{eqn:AdiabaticSimilar}
\end{multline}
This is equivalent to~(\ref{eqn:EternalAdiabaticLimit}).

Now, if $B$ and $C$ are physical generators, the spectrum of $\gamma B+C$ is confined in the left half-plane (the real parts of the eigenvalues are nonpositive), and purely imaginary eigenvalues are semisimple (the corresponding eigenspaces are diagonalizable and have no nilpotents).
Due to the similarity~(\ref{eqn:Similarity}), the adiabatic generator $\gamma B+D$ has the same spectrum as $\gamma B+C$.
Therefore, $\rme^{t(\gamma B+D)}$, as well as $\rme^{t(\gamma B+C)}$, are bounded semigroups, i.e.,
\begin{equation}
\|\rme^{t(\gamma B+C)}\| \le M, \quad 
\|\rme^{t(\gamma B+D)}\|\le M,
\label{eq:Mdef}
\end{equation}
for some $M\ge1$ for all $t\ge0$ and $\gamma\ge0$.
This ensures that the distance between the true evolution $\rme^{t(\gamma B+C)}$ and the adiabatic approximation $\rme^{t(\gamma B+D)}$, namely the norm of~(\ref{eqn:AdiabaticSimilar}), is bounded by 
\begin{equation}
 \bigl\|\rme^{t(\gamma B+C)}-\rme^{t(\gamma B+D)} \bigr\|
\le\frac{2M}{\gamma}\sum_\ell\|S_\ell\Omega_\ell P_\ell \|,
\label{eqn:EternalAdiabatic}
\end{equation}
for all $t\ge0$.
This means that the adiabatic evolution $\rme^{t(\gamma B+D)}$ is a good approximation to the true evolution $\rme^{t(\gamma B+C)}$ with the error remaining $O(1/\gamma)$ \textit{for all times $t\ge0$}.
This proves the \textit{eternal adiabaticity} of the evolution, and this is the central result of this paper.

In the norm induced by the operator trace norm, we have $\|\rme^{t(\gamma B+C)}\|=1$ for a physical evolution~\cite{ref:PrezGarcaWolfPetzRuskai-JMP2006}, and the distance~(\ref{eqn:EternalAdiabatic}) can be explicitly bounded by
\begin{equation}
\bigl\|\rme^{t(\gamma B+C)}-\rme^{t(\gamma B+D)}\bigr\|
< \frac{1}{\gamma}
\sum_\ell \gamma_\ell \|P_\ell\|,
\label{eqn:EternalBoundD}
\end{equation}
for $\gamma\ge 2 \max_\ell \gamma_\ell$,
where
\begin{equation}
\gamma_\ell = 4 \|S_\ell\|\|C\|\|P_\ell\| \frac{1-(\|S_\ell\|\|N_\ell\|)^{n_\ell}}{1-\|S_\ell\|\|N_\ell\|}.
\label{eq:gammaellt}
\end{equation}
See Appendix~\ref{app:EternalBounds} for its derivation and its tighter bound valid also for other norms.

Note that in the unitary case $\|P_\ell\|=1$, $\|N_\ell\|=0$ and hence $\gamma_\ell=4\|S_\ell\|\|C\| \le4\|C\|/ \eta$, where $\eta$ is the spectral gap of $B$.

\section{Conjugate Adiabatic Bloch Equation}
\label{sec:ConjugateBloch}
One might have noticed the asymmetry in the perturbative expressions~(\ref{eqn:D2})--(\ref{eqn:D3}) for the second- and higher-order terms.
This asymmetry stems from the asymmetry in the derivation of the adiabatic theorem.
We can think of an alternative way of estimating the difference between an adiabatic evolution and the true evolution.
Instead of~(\ref{eqn:difference0}), we can proceed as
\begin{align}
&
P_\ell
\bigl(
\rme^{t(\gamma B+C)}
-\rme^{t(\gamma B+D_\ell)}
\bigr)
\nonumber\\
&\quad
=-
P_\ell
\int_0^t\rmd s\,\frac{\partial }{\partial s} \bigl(
\rme^{s(\gamma B+D_\ell)}
\rme^{(t-s)(\gamma B+C)}
\bigr)
\nonumber
\displaybreak[0]\\
&\quad
=\int_0^t\rmd s\,
\rme^{s(\gamma B+D_\ell )}
P_\ell
(C-D_\ell )
\rme^{(t-s)(\gamma B+C)}.
\label{eqn:difference1}
\end{align}
Notice the difference in the order of the operators compared to~(\ref{eqn:difference0}).
The components are the same but they are ordered in the opposite order.
We can repeat the same steps followed above, starting from this reverted expression~(\ref{eqn:difference1}).
We can derive an adiabatic theorem, we can iteratively apply the adiabatic theorem to improve the adiabatic approximation, and we can prove the eternal adiabaticity.
All the formulas originating from~(\ref{eqn:difference1}) are similar to those obtained above, but the orders of operators are exactly reverted.

Let us collect the main formulas.
We get a new set of adiabatic Bloch equations
\begin{equation}
\frac{1}{\gamma}\tilde{\Omega}_\ell^2S_\ell-\tilde{\Omega}_\ell\left(1+\frac{1}{\gamma}S_\ell C\right)
	 +N_\ell\tilde{\Omega}_\ell S_\ell+P_\ell C
=0,
\label{eqn:AdiabaticBlochRev}
\end{equation}
with 
\begin{equation}
(1-P_\ell)\tilde{\Omega}_\ell =0 ,
\label{eqn:AB2Rev}
\end{equation}
from the iterated adiabatic theorem based on the reversed equation~(\ref{eqn:difference1}).
Compare them with~(\ref{eqn:AdiabaticBloch}) and~(\ref{eqn:AB2}).
Now, by choosing as eternal adiabatic generator 
\begin{equation}
\tilde{D}_\ell=\tilde{\Omega}_\ell P_\ell = P_\ell\tilde{\Omega}_\ell P_\ell,
\label{eqn:Omega2DRev}
\end{equation}
we get
\begin{multline}
\rme^{t
(\gamma B+C)
}
-\rme^{t(\gamma B+\tilde{D})}
\\
=
\frac{1}{\gamma}
\sum_\ell
\bigl(
P_\ell \tilde{\Omega}_\ell
S_\ell
\rme^{t(\gamma B+C)}
-
\rme^{t(\gamma B+\tilde{D})}
P_\ell \tilde{\Omega}_\ell
S_\ell
\bigr),
\label{eqn:AdiabaticSimilarRev}
\end{multline}
where
\begin{equation}
\tilde{D}=\sum_\ell\tilde{D}_\ell.
\end{equation}
This is the counterpart of~(\ref{eqn:AdiabaticSimilar}).
The similarity between $\gamma B+\tilde{D}$ and $\gamma B+C$ also holds.
We have the intertwining relations
\begin{equation}
\tilde{U}_\ell(\gamma B+C)
=(\gamma B+\tilde{D}_\ell)\tilde{U}_\ell,
\label{eqn:UD=HURev}
\end{equation}
for
\begin{equation}
\tilde{U}_\ell=P_\ell-\frac{1}{\gamma}\tilde{\Omega}_\ell S_\ell,
\label{eqn:UellRev}
\end{equation}
and the similarity relation
\begin{equation}
\gamma B+\tilde{D}
=
\tilde{U}(\gamma B+C)\tilde{U}^{-1},
\end{equation}
with
\begin{equation}
\tilde{U}=\sum_\ell\tilde{U}_\ell
=1-\frac{1}{\gamma}\sum_\ell\tilde{\Omega}_\ell S_\ell.
\label{Utilde}
\end{equation}
These correspond to~(\ref{eqn:UD=HU}) and~(\ref{eqn:Similarity}), respectively.
Note that $\tilde{U}_\ell$ satisfies
\begin{equation}
P_\ell\tilde{U}_\ell=\tilde{U}_\ell,\qquad
\tilde{U}_\ell P_\ell=P_\ell,
\label{eqn:UPPURev}
\end{equation}
similarly to~(\ref{eqn:UPPU}).
The equation for $\tilde{U}_\ell$ is given by
\begin{equation}
\tilde{U}_\ell
-N_\ell\tilde{U}_\ell S_\ell
+\frac{1}{\gamma}(\tilde{U}_\ell C-\tilde{U}_\ell C\tilde{U}_\ell)S_\ell
-P_\ell=0.
\label{eqn:BlochRev}
\end{equation}
Compare it with~(\ref{eqn:Bloch}).

In the unitary case, $C$ and $S_\ell$ are skew-Hermitian, $P_\ell$ is Hermitian, and there is no nilpotent $N_\ell$.
Comparing the Bloch equation for $\Omega_\ell$ in~(\ref{eqn:AdiabaticBloch}) and the one for $\tilde{\Omega}_\ell$ in~(\ref{eqn:AdiabaticBlochRev}), one realizes that $\tilde{\Omega}_\ell=-\Omega_\ell^\dag$, and hence, $\tilde{U}_\ell=U_\ell^\dag$.
This alternative approach is therefore  a conjugate version of the original approach in the unitary case.

\section{Generalized Schrieffer-Wolff Transformation for Open Systems}
\label{sec:SW}
In the unitary case, where $B$ and $C$ are both skew-Hermitian with no nilpotent in $B$, the asymmetry in the perturbative expressions~(\ref{eqn:D2})--(\ref{eqn:D3}) leads to a non-skew-Hermitian $D$, in spite of the skew-Hermiticity of $B$ and $C$\@.
This fact is known in the literature~\cite{Bloch,Cloizeaux,Lindgren,Klein,Bravyi,Sanz}.
This does not spoil the validity of the approximation and the eternal adiabaticity, but it would be nicer if we could have an effective generator that has the correct structure as a physical generator (i.e.~skew-Hermitian in the unitary case) and works equally well as $D$ as an approximation.

In the unitary case, it is known that the perturbative series~(\ref{eqn:D0})--(\ref{eqn:D3}) can be made symmetric and the skew-Hermiticity of the adiabatic generator $D$ can be amended via an additional similarity transformation~\cite{Bloch,Cloizeaux}.
We can generalize it for open systems.
It provides us with a generalization of the Schrieffer-Wolff transformation~\cite{Schrieffer,Shavitt,Bravyi} for open systems~\cite{Kessler}.

Let us first show that
\begin{equation}
\tilde{P}_\ell
=U_\ell(\tilde{U}_\ell U_\ell)^{-1}\tilde{U}_\ell
\label{eqn:ProjectionPerturbed}
\end{equation}
is the projection onto the direct sum of the eigenspaces of $\gamma B+C$ belonging to the eigenvalues perturbed from the unperturbed eigenvalue $\gamma b_\ell$ of $\gamma B$.
Here, $(\tilde{U}_\ell U_\ell)^{-1}$ is the inverse of $\tilde{U}_\ell U_\ell$ on $P_\ell$, defined by
\begin{equation}
(\tilde{U}_\ell U_\ell)^{-1}
=\left(
1 +\frac{1}{\gamma^2}\tilde{\Omega}_\ell S_\ell^2\Omega_\ell
\right)^{-1}
P_\ell.
\label{eqn:pseudoinv}
\end{equation}
Note the properties $\Omega_\ell=\Omega_\ell P_\ell$ in~(\ref{eqn:AB2}), $\tilde{\Omega}_\ell=P_\ell\tilde{\Omega}_\ell$ in~(\ref{eqn:AB2Rev}), $U_\ell P_\ell=U_\ell$, $P_\ell U_\ell=P_\ell$ in~(\ref{eqn:UPPU}), and $P_\ell\tilde{U}_\ell=\tilde{U}_\ell$, $\tilde{U}_\ell P_\ell=P_\ell$ in~(\ref{eqn:UPPURev}).
Thus
\begin{equation}
\tilde{U}_\ell U_\ell = P_\ell \tilde{U}_\ell U_\ell P_\ell,    \quad (\tilde{U}_\ell U_\ell)^{-1} = P_\ell (\tilde{U}_\ell U_\ell)^{-1} P_\ell
\end{equation} 
reside in the subspace $P_\ell$.
Now, $\tilde{P}_\ell$ is clearly a projection, satisfying $\tilde{P}_\ell^2=\tilde{P}_\ell$.
In addition, $\tilde{P}_\ell$ commutes with $\gamma B+C$.
Indeed,
\begin{align}
(\gamma B+C)\tilde{P}_\ell
&=
(\gamma B+C)U_\ell(\tilde{U}_\ell U_\ell)^{-1}\tilde{U}_\ell
\nonumber\\
&=
U_\ell(\gamma B+D_\ell)(\tilde{U}_\ell U_\ell)^{-1}\tilde{U}_\ell
\nonumber\\
&=
U_\ell(\tilde{U}_\ell U_\ell)^{-1}(\gamma B+\tilde{D}_\ell)\tilde{U}_\ell
\nonumber\\
&=
U_\ell(\tilde{U}_\ell U_\ell)^{-1}\tilde{U}_\ell(\gamma B+C)
\nonumber\\
&=
\tilde{P}_\ell(\gamma B+C),
\end{align}
where we have used the intertwining relations~(\ref{eqn:UD=HU}) and~(\ref{eqn:UD=HURev}) for the second and fourth equalities, respectively, and for the third equality we have used
\begin{equation}
(\gamma B+D_\ell)(\tilde{U}_\ell U_\ell)^{-1}
=
(\tilde{U}_\ell U_\ell)^{-1}(\gamma B+\tilde{D}_\ell),
\end{equation}
which follows from 
\begin{equation}
\tilde{U}_\ell U_\ell(\gamma B+D_\ell)
=
\tilde{U}_\ell(\gamma B+C)U_\ell
=
(\gamma B+\tilde{D}_\ell)\tilde{U}_\ell U_\ell.
\end{equation}
Observe also that $\tilde{P}_\ell\to P_\ell$ as $\gamma\to+\infty$, and the eigenvalues of $\tilde{P}_\ell(\gamma B+C)\tilde{P}_\ell$ are close to $\gamma b_\ell$ for large $\gamma$.
These facts imply that $\tilde{P}_\ell$ is the projection onto the direct sum of the eigenspaces of $\gamma B+C$ corresponding to the eigenprojection $P_\ell$ of $B$.

In Ref.~\cite{Bravyi}, it is pointed out that the Schrieffer-Wolff transformation for the unitary case is nothing but the ``direct rotation'' $(\tilde{P}_\ell P_\ell)^{1/2}$ connecting $P_\ell$ and $\tilde{P}_\ell$~\cite[Definition 2.2]{Bravyi}.
A natural generalization of the Schrieffer-Wolff transformation for open systems, namely, a natural generalization of the direct rotation, is thus provided by
\begin{equation}
W_\ell
=(\tilde{P}_\ell P_\ell)^{1/2}
=U_\ell(\tilde{U}_\ell U_\ell)^{-1/2},
\label{eqn:SW}
\end{equation}
where $(\tilde{U}_\ell U_\ell)^{-1/2}$ is the square root of $(\tilde{U}_\ell U_\ell)^{-1}$ defined in~(\ref{eqn:pseudoinv}).
We use the primary square root such that $(\tilde{P}_\ell P_\ell)^{1/2}\to P_\ell$ and $(\tilde{U}_\ell U_\ell)^{-1/2}\to P_\ell$ in the limit $\gamma\to+\infty$ (see e.g.\ Refs.~\cite[Chap.~1]{ref:MatrixFunctions-Higham} and~\cite[Sec.~6.4]{ref:MatrixAnalysisTopics-HornJohnson} for primary matrix function).
The equivalence of the last two expressions in~(\ref{eqn:SW}) can be verified by looking at their squares, $U_\ell(\tilde{U}_\ell U_\ell)^{-1/2}U_\ell(\tilde{U}_\ell U_\ell)^{-1/2}=U_\ell(\tilde{U}_\ell U_\ell)^{-1}=U_\ell(\tilde{U}_\ell U_\ell)^{-1}\tilde{U}_\ell P_\ell=\tilde{P}_\ell P_\ell$, where we have used $P_\ell U_\ell=P_\ell$ and $\tilde{U}_\ell P_\ell=P_\ell$.
This $W_\ell$ connects $P_\ell$ and $\tilde{P}_\ell$ as
\begin{equation}
W_\ell
=W_\ell P_\ell
=
\tilde{P}_\ell
W_\ell,
\label{eqn:IntertwiningW}
\end{equation}
which can be verified trivially on the basis of the definitions of $\tilde{P}_\ell$ and $W_\ell$ in~(\ref{eqn:ProjectionPerturbed}) and~(\ref{eqn:SW}), respectively.
Then, 
\begin{equation}
\gamma B_\ell+K_\ell=W_\ell^{-1}(\gamma B+C)W_\ell
\label{eqn:SWHamiltonian}
\end{equation}
provides an effective generator which has the same block structure as $B$, where $W_\ell^{-1}$ is a pseudoinverse satisfying
\begin{equation}
W_\ell^{-1}W_\ell=P_\ell,\qquad
W_\ell W_\ell^{-1}=\tilde{P}_\ell,
\end{equation}
which is explicitly given by
\begin{equation}
W_\ell^{-1}
=
(P_\ell\tilde{P}_\ell)^{1/2}
=
(\tilde{U}_\ell U_\ell)^{-1/2}\tilde{U}_\ell.
\label{eqn:SWinv}
\end{equation}
This $W_\ell^{-1}$ brings $\tilde{P}_\ell$ back to $P_\ell$ as
\begin{equation}
W_\ell^{-1}\tilde{P}_\ell
=P_\ell W_\ell^{-1}
=W_\ell^{-1}.
\label{eqn:IntertwiningWinv}
\end{equation}

In the unitary case, $P_\ell=P_\ell^\dag$ and $\tilde{U}_\ell=U_\ell^\dag$ (see Sec.~\ref{sec:ConjugateBloch}), and the polar decomposition of $U_\ell$ reads $U_\ell=V_\ell|U_\ell|$, where $|U_\ell|=(U_\ell^\dag U_\ell)^{1/2}$ and $V_\ell$ is some unitary.
Thus, in the unitary case, $W_\ell$ in~(\ref{eqn:SW}) and $W_\ell^{-1}$ in~(\ref{eqn:SWinv}) are reduced to $W_\ell=V_\ell P_\ell$ and $W_\ell^{-1}=P_\ell V_\ell^\dag$, respectively, and~(\ref{eqn:SWHamiltonian}) reads
\begin{equation}
\gamma B_\ell+K_\ell=P_\ell V_\ell^\dag (\gamma B+C)V_\ell P_\ell,
\label{eqn:SWHamiltonian2}
\end{equation}
so that $K_\ell$ is guaranteed to be skew-Hermitian.
This reproduces the Schrieffer-Wolff formalism~\cite[Definition 3.1]{Bravyi}, and the transformation~(\ref{eqn:SWHamiltonian}) is a generalization of the Schrieffer-Wolff transformation for open systems.

Recalling the intertwining relations~(\ref{eqn:UD=HU}) and~(\ref{eqn:UD=HURev}), we can rewrite the Schrieffer-Wolff transformation~(\ref{eqn:SWHamiltonian}) as
\begin{align}
\gamma B_\ell+K_\ell
&=(\tilde{U}_\ell U_\ell)^{-1/2}\tilde{U}_\ell
(\gamma B+C)
U_\ell(\tilde{U}_\ell U_\ell)^{-1/2}
\nonumber\\
&=(\tilde{U}_\ell U_\ell)^{1/2}
(\gamma B+D_\ell)
(\tilde{U}_\ell U_\ell)^{-1/2}
\nonumber\displaybreak[0]\\
&=(\tilde{U}_\ell U_\ell)^{-1/2}
(\gamma B+\tilde{D}_\ell)
(\tilde{U}_\ell U_\ell)^{1/2}.
\label{eqn:SWD}
\end{align}
It is clear from the first expression of~(\ref{eqn:SWD}) that the perturbative series of $K_\ell=\sum_{j=0}^\infty K_\ell^{(j)}/\gamma^j$ is symmetric also for open systems.
The first few orders are given by
\begin{align}
K_\ell^{(0)}
={}&
P_\ell CP_\ell,
\label{eqn:SW0}
\displaybreak[0]
\\
K_\ell^{(1)}
={}&
{-\frac{1}{2}}
P_\ell
CS_\ell\overrightarrow{\langle C\rangle}
P_\ell
-
\frac{1}{2}P_\ell\overleftarrow{\langle C\rangle}S_\ell CP_\ell,
\label{eqn:SW1}
\displaybreak[0]
\\
K_\ell^{(2)}
={}&
\frac{1}{2}
P_\ell CS_\ell
\overrightarrow{
\langle
CS_\ell\langle C\rangle
\rangle
}
P_\ell
+\frac{1}{2}
P_\ell
\overleftarrow{
\langle
\langle C\rangle
S_\ell
C
\rangle
}
S_\ell CP_\ell
\nonumber
\\
&{}
-
\frac{1}{2}
P_\ell CS_\ell^2
\overrightarrow{
\langle
\langle C\rangle P_\ell C
\rangle
}
P_\ell
-
\frac{1}{2}
P_\ell
\overleftarrow{
\langle
C
P_\ell
\langle C\rangle
\rangle
}
S_\ell^2CP_\ell,
\label{eqn:SW2}
\end{align}
\begin{widetext}
\begin{align}
K_\ell^{(3)}
={}&
{
-\frac{1}{2}
P_\ell
C
S_\ell
\overrightarrow{
\langle
CS_\ell
\langle
CS_\ell\langle C\rangle
\rangle
\rangle
}
P_\ell
}
-
\frac{1}{2}
P_\ell
\overleftarrow{
\langle
\langle
\langle C\rangle
S_\ell
C
\rangle
S_\ell
C
\rangle
}
S_\ell
C
P_\ell
\nonumber
\displaybreak[0]\\
&{}
+
\frac{1}{2}
P_\ell
C
S_\ell
\overrightarrow{
\langle
CS_\ell^2
\langle
\langle C\rangle P_\ell C
\rangle
\rangle
}
P_\ell
+
\frac{1}{2}
P_\ell
\overleftarrow{
\langle
\langle
C
P_\ell
\langle C\rangle
\rangle
S_\ell^2
C
\rangle
}
S_\ell
C
P_\ell
\nonumber\\
&{}
+
\frac{1}{2}
P_\ell
C
S_\ell^2
\overrightarrow{
\langle
\langle C\rangle P_\ell
CS_\ell\langle C\rangle
\rangle
}
P_\ell
+
\frac{1}{2}
P_\ell
\overleftarrow{
\langle
\langle C\rangle
S_\ell
C
P_\ell
\langle C\rangle
\rangle
}
S_\ell^2
C
P_\ell
\nonumber
\displaybreak[0]
\\
&{}
+
\frac{1}{2}
P_\ell
C
S_\ell^2
\overrightarrow{
\langle
\langle
CS_\ell\langle C\rangle
\rangle
P_\ell
C
\rangle
}
P_\ell
+
\frac{1}{2}
P_\ell
\overleftarrow{
\langle
C
P_\ell
\langle
\langle C\rangle
S_\ell
C
\rangle
\rangle
}
S_\ell^2
C
P_\ell
\nonumber
\displaybreak[0]
\\
&{}
-
\frac{1}{2}
P_\ell
C
S_\ell^3
\overrightarrow{
\langle
\langle
\langle C\rangle P_\ell C
\rangle
P_\ell
C
\rangle
}
P_\ell
-
\frac{1}{2}
P_\ell
\overleftarrow{
\langle
C
P_\ell
\langle
C
P_\ell
\langle C\rangle
\rangle
\rangle
}
S_\ell^3
C
P_\ell
\nonumber
\displaybreak[0]
\\
&{}
-
\frac{1}{8}
N_\ell
\overleftarrow{
\langle C\rangle
}
S_\ell^2
\overrightarrow{
\langle C\rangle
}
P_\ell
\overleftarrow{
\langle C\rangle
}
S_\ell^2
\overrightarrow{
\langle C\rangle
}
P_\ell
-
\frac{1}{8}
P_\ell
\overleftarrow{
\langle C\rangle
}
S_\ell^2
\overrightarrow{
\langle C\rangle
}
P_\ell
\overleftarrow{
\langle C\rangle
}
S_\ell^2
\overrightarrow{
\langle C\rangle
}
N_\ell
\nonumber
\displaybreak[0]
\\
&{}
+
\frac{1}{4}
P_\ell
\overleftarrow{
\langle C\rangle
}
S_\ell^2
\overrightarrow{
\langle C\rangle
}
N_\ell
\overleftarrow{
\langle C\rangle
}
S_\ell^2
\overrightarrow{
\langle C\rangle
}P_\ell
,
\label{eqn:SW3}
\end{align}
\end{widetext}
where
\begin{equation}
\overrightarrow{\langle A\rangle}=\sum_{n=0}^{n_\ell-1}S_\ell^nAN_\ell^n,\quad
\overleftarrow{\langle A\rangle}=\sum_{n=0}^{n_\ell-1}N_\ell^nAS_\ell^n.
\end{equation}
The first bracket $\overrightarrow{\langle A\rangle}$ is the same as the one introduced in~(\ref{eqn:Sandwitch}), but an arrow is put here to stress the order of the operators.
Concatenated brackets like $\overrightarrow{\langle CS_\ell\overrightarrow{\langle CS_\ell\overrightarrow{\langle C\rangle}\rangle}\rangle}$ are simply denoted with a single arrow like $\overrightarrow{\langle CS_\ell\langle CS_\ell\langle C\rangle\rangle\rangle}$.
Concatenation of brackets with different orientations of arrows does not appear.
In the unitary case, this series reduces to the perturbative series obtained in Refs.~\cite{Cloizeaux,Klein}.

The generators $\gamma B+C$, $\gamma B+D$, $\gamma B+\tilde{D}$, and $\gamma B+K$ with 
\begin{equation}
K=\sum_\ell K_\ell
\end{equation}
are similar to each other, and they share the same spectrum,
\begin{align}
\gamma B+C
&=U(\gamma B+D)U^{-1}
\nonumber\displaybreak[0]\\
&=\tilde{U}^{-1}(\gamma B+\tilde{D})\tilde{U}
\nonumber\displaybreak[0]\\
&=W(\gamma B+K)W^{-1},
\label{eqn:SimilarK}
\end{align}
where $U=\sum_\ell U_\ell$ and $\tilde{U}=\sum_\ell\tilde{U}_\ell$ are introduced in~(\ref{U}) and~(\ref{Utilde}), respectively, and
\begin{equation}
W=\sum_\ell W_\ell,\quad
W^{-1}=\sum_\ell W_\ell^{-1}.
\end{equation}
Thanks to the similarity relation and its closeness to the identity $W-1= O(1/\gamma)$, the distance between the approximate adiabatic evolution $\rme^{t(\gamma B+K)}$ and the true evolution $\rme^{t(\gamma B+C)}$ remains $O(1/\gamma)$ eternally.
In the norm induced by the operator trace norm, we have $\|\rme^{t(\gamma B+C)}\|=1$ for the physical evolution \cite{ref:PrezGarcaWolfPetzRuskai-JMP2006}, and the distance can be bounded in the same way as the one for $\rme^{t(\gamma B+D)}$ given in (\ref{eqn:EternalBoundD}). That is,
\begin{equation}
\bigl\|\rme^{t(\gamma B+C)}-\rme^{t(\gamma B+K)}\bigr\|
< \frac{1}{\gamma}
\sum_\ell \gamma_\ell \|P_\ell\|,
\label{eqn:EternalBoundK}
\end{equation}
for $\gamma\geq 2\max_\ell \gamma_\ell$, 
with $\gamma_\ell$ defined in~\eqref{eq:gammaellt}. 
See Appendix~\ref{app:EternalBounds} for its derivation and its tighter bound valid also for other norms.

\section{Physical properties of the Adiabatic Generators $D$, $\tilde{D}$, and $K$}
\label{sec:Physicality}
As already mentioned, the adiabatic generator $D$ is generally not skew-Hermitian even for unitary evolution with skew-Hermitian generators $B$ and $C$.
This is easily anticipated from the asymmetry in the perturbative series in~(\ref{eqn:D0})--(\ref{eqn:D3}).
This asymmetry can be fixed by the transformation discussed in the previous section.
The adiabatic generator $K$ obtained by the generalized Schrieffer-Wolff transformation is symmetric, and it is guaranteed to be skew-Hermitian for unitary evolution.

In the nonunitary case, the structure of a physical generator is much more subtle than in the unitary case~\cite{ref:DynamicalMap-Alicki,ref:GKLS-DariuszSaverio}.
It should be Hermiticity-preserving (HP), trace-preserving (TP), and conditionally completely positive (CP) (with a positive-semidefinite Kossakowski matrix)~\cite{ref:VectorizationHavel} as a generator acting on density operators.
These impose a delicate structure on the generator, leading to the Gorini-Kossakowski-Lindbald-Sudarshan (GKLS) form~\cite{ref:DynamicalMap-Alicki,ref:GKLS-DariuszSaverio}.

In this section, we are going to show that $D$, $\tilde{D}$, and $K$ obtained for physical (i.e.~HP, TP, and CP) generators $B$ and $C$ acting on density operators are both HP and TP in the general nonunitary case (including the unitary case).
On the other hand, CP is not guaranteed in the nonunitary case, even for the symmetric $K$, as we will see in the next section.

\subsection{$D$, $\tilde{D}$, and $K$ are TP}
Note first that the spectrum $\{b_\ell\}$ of a physical generator $B$ acting on density operators is contained in the closed left half-plane $\Re b_\ell\le0$, and $B$ always has $b_0=0$ in its spectrum.
In addition, purely imaginary eigenvalues $b_\ell\in\rmi\mathbb{R}$ including $b_0=0$ are semisimple, that is $P_\ell B P_\ell = b_\ell P_\ell$ are diagonalizable with no nilpotents.
See e.g.~Ref.~\cite{ref:Mixing-Wolf,ref:TextbookWatrous}, in particular Propositions 6.1--6.3 and Theorem 6.1 of Ref.~\cite{ref:Mixing-Wolf}.

Since $B$ is assumed to be a physical generator, it is TP, i.e., $\tr[B(\sigma)]=0$ for any operators $\sigma$ acting on the Hilbert space.
Since this can be written as $\tr[B(\sigma)]=(\openone|B(\sigma))=(\openone|B|\sigma)=0$, with $(\varrho|\sigma)=\tr(\varrho^\dag\sigma)$ being the Hilbert-Schmidt inner product of operators $\varrho$ and $\sigma$ acting on the Hilbert space, the TP of $B$ as a generator is represented by
\begin{equation}
(\openone|B=0.
\end{equation}
Projecting it by $P_\ell$ from the right, we get
\begin{equation}
(\openone|BP_\ell
=(\openone|(b_\ell P_\ell+N_\ell)
=0.
\label{eqn:TPell}
\end{equation}
This condition is trivial for $\ell=0$, since $b_0=0$ and there is no nilpotent $N_0=0$ in this sector.
For nonvanishing eigenvalues $b_\ell$, let us multiply $N_\ell^{n_\ell-1}$ from the right of~(\ref{eqn:TPell}).
It yields $(\openone|N_\ell^{n_\ell-1}=0$, since $N_\ell^{n_\ell}=0$, $P_\ell N_\ell=N_\ell$, and $b_\ell\neq0$.
Then, by multiplying $N_\ell^{n_\ell-2}$ from the right of ~(\ref{eqn:TPell}) again, we realize that $(\openone|N_\ell^{n_\ell-2}=0$.
After $n_\ell-1$ such iterations, we reach
\begin{equation}
(\openone|N_\ell=0.
\label{eqn:TrN}
\end{equation}
This further implies
\begin{equation}
(\openone|P_\ell=0\quad\text{for}\quad b_\ell\neq0.
\end{equation}
Finally, since $\sum_\ell P_\ell=1$, we need to have
\begin{equation}
(\openone|P_0=(\openone|,
\label{eqn:P0TP}
\end{equation}
namely, $P_0$ too is TP\@.

Now, let us look at the adiabatic Bloch equation~(\ref{eqn:AdiabaticBlochRev}) for $\tilde{\Omega}_\ell$.
Putting $(\openone|$ on the left of the adiabatic Bloch equation, we get
\begin{equation}
(\openone|\tilde{\Omega}_\ell\left(
1+\frac{1}{\gamma}S_\ell C
-\frac{1}{\gamma}\tilde{\Omega}_\ell S_\ell
\right)
=0,
\end{equation}
where we have used~(\ref{eqn:TrN})--(\ref{eqn:P0TP}) and $(\openone|C=0$.
This implies
\begin{equation}
(\openone|\tilde{\Omega}_\ell=0
\end{equation}
for large enough $\gamma$, since $1+\frac{1}{\gamma}S_\ell C
-\frac{1}{\gamma}\tilde{\Omega}_\ell S_\ell$ is invertible.
Therefore, we have
\begin{equation}
(\openone|\tilde{U}_\ell
=
(\openone|\left(
P_\ell-\frac{1}{\gamma}\tilde{\Omega}_\ell S_\ell
\right)
=
(\openone|P_\ell,
\end{equation}
and
\begin{equation}
(\openone|(\tilde{U}_\ell U_\ell)^\alpha
=
(\openone|\left(
1+\frac{1}{\gamma^2}\tilde{\Omega}_\ell S_\ell^2\Omega_\ell
\right)^\alpha
P_\ell
=
(\openone|P_\ell
\end{equation}
for $\alpha=-1$ and $-1/2$.
Recall the definition of the pseudoinverse $(\tilde{U}_\ell U_\ell)^{-1}$ in~(\ref{eqn:pseudoinv}).
Then, it immediately follows that $D$, $\tilde{D}$, and $K$ are TP\@.
For instance, using the similarity in~(\ref{eqn:SWD}), the adiabatic generator $D$ is proved to be TP as
\begin{align}
(\openone|D
&=\sum_\ell
(\openone|[
(\tilde{U}_\ell U_\ell)^{-1}\tilde{U}_\ell(\gamma B+C)U_\ell
-\gamma B_\ell
]
\nonumber
\displaybreak[0]
\\
&=
(\openone|P_0CU_0
\vphantom{\sum_\ell}
\nonumber
\displaybreak[0]
\\
&=0.
\vphantom{\sum_\ell}
\end{align}
TP of $\tilde{D}$ and $K$ can be proved in the same way.

\subsection{$D$, $\tilde{D}$, and $K$ are HP}
Let us next prove that $D$, $\tilde{D}$, and $K$ are HP\@.
To this end, it is convenient to introduce an orthogonal basis of Hermitian matrices $\{\tau_0,\tau_1,\ldots\tau_{d^2-1}\}$ for a $d$-dimensional system.
Here, $\tau_0=\openone$ is the $d\times d$ identity matrix, and the $d\times d$ matrices $\tau_i$ ($i=1,\ldots,d^2-1$) are Hermitian $\tau_i=\tau_i^\dag$ and traceless $\tr\tau_i=0$, which are orthogonal to each other with respect to the Hilbert-Schmidt inner product, $(\tau_i|\tau_j)=\tr(\tau_i^\dag\tau_j)=2\delta_{ij}$ ($i,j=1,\ldots,d^2-1$).
The matrix representation $\mathsf{B}_{ij}=(\tau_i|B|\tau_j)=(\tau_i|B(\tau_j))$ ($i,j=0,1,\ldots,d^2-1$) of $B$ in such a basis is the generator of the evolution of the coherence vector $r_i=(\tau_i|\varrho)$ ($i=0,1,\ldots,d^2-1$) representing the density operator $\varrho$ of the system.
Notice that the coherence vector $(r_0,r_1,\ldots r_{d^2-1})$ corresponding to a Hermitian density operator $\varrho$ is a real vector.
Therefore, the matrix elements $\mathsf{B}_{ij}$  
of a physical generator $B$ should be all real, since $B$ should preserve the Hermiticity of density operator $\varrho$ and hence the reality of the coherence vector.
In other words, the reality of $\mathsf{B}_{ij}$ is equivalent to HP of $B$\@.
Let us call the spectral projections and nilpotents of the real matrix $\mathsf{B}$ in this representation $\mathsf{P}_\ell$ and $\mathsf{N}_\ell$, respectively.

We note that all the nonreal eigenvalues of a real matrix occur in conjugate pairs.
In addition, the spectral projections and the nilpotents of the real matrix $\mathsf{B}=\mathsf{B}^*$ satisfy
\begin{gather}
\mathsf{P}_\ell=\mathsf{P}_{\bar{\ell}}^*,\qquad
\mathsf{N}_\ell=\mathsf{N}_{\bar{\ell}}^*,
\label{eqn:PNconj}
\end{gather}
where $*$ of a matrix represents the elementwise complex conjugation and $\bar{\ell}$ refers to its complex conjugate eigenvalue, $b_{\bar{\ell}}=b_\ell^*$.
Indeed, 
  the spectral projection $\mathsf{P}_\ell$ can be constructed by 
\begin{equation}
\mathsf{P}_\ell=\int_{\mathcal{C}_\ell}\frac{\rmd z}{2\pi\rmi}(z-\mathsf{B})^{-1},
\end{equation}
where 
$\mathcal{C}_\ell$ is a contour  running anticlockwise around the eigenvalue $b_\ell$ on the complex $z$ plane \cite{ref:KatoBook}.
Since $\mathsf{B}=\mathsf{B}^*$ is real and $\mathcal{C}_\ell$ is flipped to $-\mathcal{C}_{\bar{\ell}}$ (running clockwise around the complex conjugate eigenvalue $b_\ell^*=b_{\bar{\ell}}$) by  complex conjugation, we get $\mathsf{P}_\ell^*=-\int_{-\mathcal{C}_{\bar{\ell}}}\frac{\rmd z}{2\pi\rmi}(z-\mathsf{B}^*)^{-1}=\int_{\mathcal{C}_{\bar{\ell}}}\frac{\rmd z}{2\pi\rmi}(z-\mathsf{B})^{-1}=\mathsf{P}_{\bar{\ell}}$, and $\mathsf{N}_\ell^*=[(\mathsf{B}-b_\ell)\mathsf{P}_\ell]^*=(\mathsf{B}^*-b_\ell^*)\mathsf{P}_\ell^*=(\mathsf{B}-b_{\bar{\ell}})\mathsf{P}_{\bar{\ell}}=\mathsf{N}_{\bar{\ell}}$.
This proves~(\ref{eqn:PNconj}).
This symmetry is inherited by the reduced resolvents,
\begin{equation}
\mathsf{S}_\ell
=\sum_{k\neq\ell}(b_k-b_\ell+\mathsf{N}_k)^{-1}\mathsf{P}_k
=\mathsf{S}_{\bar{\ell}}^*.
\end{equation}
Now, let us look at the adiabatic Bloch equation~(\ref{eqn:AdiabaticBloch}) in this representation,
\begin{equation}
\frac{1}{\gamma}\mathsf{S}_\ell\mathsf{\Omega}_\ell^2-\left(\mathsf{I}+\frac{1}{\gamma}\mathsf{C}\mathsf{S}_\ell\right)\mathsf{\Omega}_\ell	 +\mathsf{S}_\ell\mathsf{\Omega}_\ell\mathsf{N}_\ell+\mathsf{C}\mathsf{P}_\ell
=0.
\label{eqn:AdiabaticBlochAffine}
\end{equation}
Note that the matrix representation $\mathsf{C}$ of $C$ is also a real matrix, since $C$ is assumed to be physical.
Taking the complex conjugation of this adiabatic Bloch equation~(\ref{eqn:AdiabaticBlochAffine}) yields
\begin{equation}
\frac{1}{\gamma}\mathsf{S}_{\bar{\ell}}{\mathsf{\Omega}_\ell^*}^2-\left(\mathsf{I}+\frac{1}{\gamma}\mathsf{C}\mathsf{S}_{\bar{\ell}}\right)\mathsf{\Omega}_\ell^*
	 +\mathsf{S}_{\bar{\ell}}\mathsf{\Omega}_\ell^*\mathsf{N}_{\bar{\ell}}+\mathsf{C}\mathsf{P}_{\bar{\ell}}
=0,
\end{equation}
which implies
\begin{equation}
\mathsf{\Omega}_\ell^*=\mathsf{\Omega}_{\bar{\ell}}.
\end{equation}
By looking at the conjugate adiabatic Bloch equation~(\ref{eqn:AdiabaticBlochRev}) for $\tilde{\Omega}_\ell$, we also confirm that $
\tilde{\mathsf{\Omega}}_\ell^*=\tilde{\mathsf{\Omega}}_{\bar{\ell}}
$.
The operators $U_\ell$ and $\tilde{U}_\ell$ are also endowed with the same symmetry, $
\mathsf{U}_\ell^*=\mathsf{U}_{\bar{\ell}}
$, $
\tilde{\mathsf{U}}_\ell^*=\tilde{\mathsf{U}}_{\bar{\ell}}
$, and so are the adiabatic generators.
For instance,
\begin{align}
\mathsf{D}_\ell^*
&=
(\tilde{\mathsf{U}}_\ell^*\mathsf{U}_\ell^*)^{-1}\tilde{\mathsf{U}}_\ell^*(\gamma\mathsf{B}^*+\mathsf{C}^*)\mathsf{U}_\ell^*
-\gamma\mathsf{B}^*\mathsf{P}_\ell^*
\nonumber
\\
&=
(\tilde{\mathsf{U}}_{\bar{\ell}}\mathsf{U}_{\bar{\ell}})^{-1}\tilde{\mathsf{U}}_{\bar{\ell}}(\gamma\mathsf{B}+\mathsf{C})\mathsf{U}_{\bar{\ell}}
-\gamma\mathsf{B}\mathsf{P}_{\bar{\ell}}
\nonumber
\\
&=\mathsf{D}_{\bar{\ell}}.
\end{align}
Therefore,
\begin{equation}
\mathsf{D}=\sum_\ell\mathsf{D}_\ell=\sum_\ell\mathsf{D}_\ell^*=\mathsf{D}^*.
\end{equation}
The reality of $\tilde{\mathsf{D}}$ and $\mathsf{K}$ can be shown in the same way, and hence, $D$, $\tilde{D}$, and $K$ are HP\@.

\section{Examples}
\label{sec:Examples}
Let us look at some examples.

\begin{figure}[b]
\includegraphics[width=0.3\textwidth]{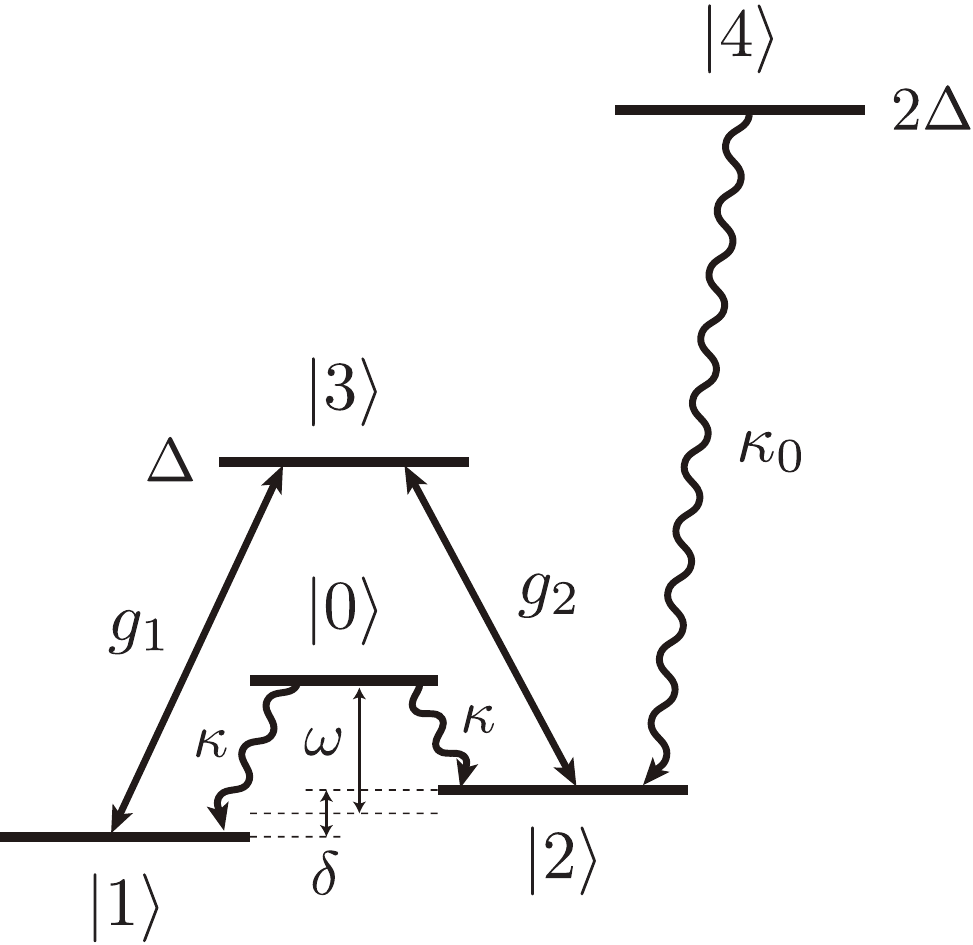}
\caption{A dissipative five-level system. Levels $\ket{1}$, $\ket{2}$, and $\ket{3}$ constitute a $\Lambda$ configuration, and there is strong decay from $\ket{4}$ to $\ket{2}$ with decay rate $\kappa_0$ and weak decay from $\ket{0}$ to $\ket{1}$ and from $\ket{0}$ to $\ket{2}$ with decay rate $\kappa$.}
\label{fig:DissipativeLambda}
\end{figure}
\subsection{Dissipative Lambda System}
\label{sec:DissipativeLambda}
We consider a five-level system, whose level structure is depicted in Fig.~\ref{fig:DissipativeLambda}.
The Hamiltonian is given by
\begin{equation}
H_\Lambda=
\begin{pmatrix}
\omega&0&0&0&0\\
0&-\delta/2&0&g_1^*/2&0\\
0&0&\delta/2&g_2^*/2&0\\
0&g_1/2&g_2/2&\Delta&0\\
0&0&0&0&2\Delta
\end{pmatrix}.
\label{eqn:5levelH}
\end{equation}
Levels $\ket{1}$, $\ket{2}$, and $\ket{3}$ constitute a $\Lambda$ configuration, and there is strong decay from $\ket{4}$ to $\ket{2}$ with decay rate $\kappa_0$ and weak decay from $\ket{0}$ to $\ket{1}$ and from $\ket{0}$ to $\ket{2}$ with decay rate $\kappa$.
We are interested in the situation where $\Delta,\kappa_0\gg\omega,|\delta|,|g_{1,2}|,\kappa$. 
Let us normalize the physical parameters $\Delta$, $\omega$, $\delta$, $g_{1,2}$, $\kappa$, and $\kappa_0$ by some unit of frequency $g_0$, and set $\gamma=\Delta/g_0$, which is considered to be much greater than $\tilde{\omega}=\omega/g_0$, $\tilde{\delta}=\delta/g_0$, $\tilde{g}_{1,2}=g_{1,2}/g_0$, $\tilde{\kappa}=\kappa/g_0$, while $\tilde{\kappa}_0=\kappa_0/\Delta= O(1)$.
We apply our formalism to Markovian generators of the GKLS form
\begin{align}
B
&=-\rmi[H_0,{}\bullet{}]
-\frac{1}{2}\tilde{\kappa}_0(
L_0^\dag L_0{}\bullet{}
+{}\bullet{}L_0^\dag L_0
-2L_0{}\bullet{}L_0^\dag
)
,
\nonumber
\displaybreak[0]
\\
C&=-\rmi[H_I,{}\bullet{}]
-\frac{1}{2}\tilde{\kappa}
\sum_{i=1,2}(
L_i^\dag L_i{}\bullet{}
+{}\bullet{}L_i^\dag L_i
-2L_i{}\bullet{}L_i^\dag
),
\label{eqn:DissipativeLambdaBC}
\end{align}
with
\begin{align}
H_0&=
\begin{pmatrix}
0&0&0&0&0\\
0&0&0&0&0\\
0&0&0&0&0\\
0&0&0&1&0\\
0&0&0&0&2
\end{pmatrix}
,
&&
L_0
=
\ket{2}\bra{4},
\nonumber\\
H_I&=
\begin{pmatrix}
\tilde{\omega}&0&0&0&0\\
0&-\tilde{\delta}/2&0&\tilde{g}_1^*/2&0\\
0&0&\tilde{\delta}/2&\tilde{g}_2^*/2&0\\
0&\tilde{g}_1/2&\tilde{g}_2/2&0&0\\
0&0&0&0&0
\end{pmatrix}
,
&&
\begin{cases}
\medskip
L_1
=
\ket{1}\bra{0},\\
L_2
=
\ket{2}\bra{0}.
\end{cases}
\label{eqn:DissipativeLambdaHL}
\end{align}
The $\Lambda$ system is a standard setup to discuss adiabatic elimination, and approximations beyond the adiabatic elimination have been studied on the platform of the $\Lambda$ system in the literature~\cite{Brion,Sanz}.
Here, we can deal with the $\Lambda$ system in the presence of noise.
By abuse of notation, we will omit tildes $\tilde{\omega}\to\omega$, $\tilde{\delta}\to\delta$, $\tilde{g}_{1,2}\to g_{1,2}$, $\tilde{\kappa}\to\kappa$, and $\tilde{\kappa}_0\to\kappa_0$ in the following analysis.

\begin{figure}
\includegraphics[width=0.45\textwidth]{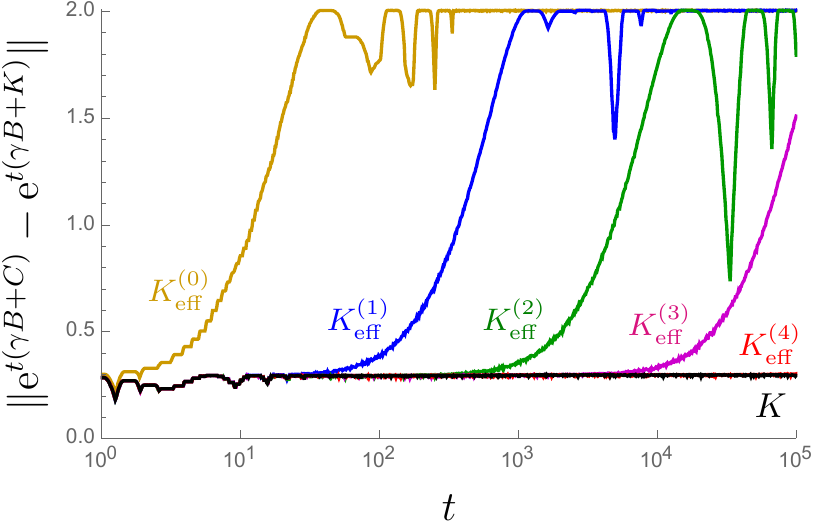}
\caption{
Operator-norm distances as functions of time $t$ between the full evolution $\rme^{t(\gamma B+C)}$ and the $k$th-order adiabatic approximations of the form $\rme^{t(\gamma B+K_\text{eff}^{(k)})}$ with $K_\text{eff}^{(k)}=\sum_{j=0}^kK^{(j)}/\gamma^j$ ($k=0,1,2,3,4,\infty$), for the dissipative 5-level system~(\ref{eqn:5levelH}) with a $\Lambda$ structure (see Fig.~\ref{fig:DissipativeLambda}).
The parameters are set at $\tilde{\delta}=\tilde{g}_1=\tilde{g}_2=1$, $\tilde{\kappa}=0.001$, $\tilde{\kappa}_0=1$, and $\gamma=10$.
We have chosen the spectral norm (the maximum of the singular values) of a matrix representation of the map to estimate the distance.
The distances actually oscillate radically as quasiperiodic functions of time: their upper envelopes are plotted here.
It is clearly observed that the $k$th-order approximation $K_\text{eff}^{(k)}$ works well for times up to $t= O(\gamma^{k+1})$, while the nonperturbative adiabatic generator $K=K_\text{eff}^{(\infty)}$ works eternally with the error remaining $O(1/\gamma)$ for long times.
}
\label{fig:DissipativeLambdaEvolution}
\end{figure}

According to the perturbative formulas in~(\ref{eqn:SW0})--(\ref{eqn:SW3}), we get the $j$th-order term $K^{(j)}=\sum_\ell K_\ell^{(j)}$ of the adiabatic generator $K=\sum_{j=0}^\infty K^{(j)}/\gamma^j$ in the GKLS form \cite{note:GKLS} 
\begin{align}
K^{(j)}
={}&{-\rmi}[H^{(j)},{}\bullet{}]
\nonumber
\\
&{}-\frac{1}{2}\sum_i\Gamma_i^{(j)}
(
L_i^{(j)\dag}L_i^{(j)}{}\bullet{}
+{}\bullet{}L_i^{(j)\dag}L_i^{(j)}
\nonumber\\[-2.2truemm]
&\qquad\qquad\qquad\qquad\qquad\ \ {}
-
2L_i^{(j)}{}\bullet{}L_i^{(j)\dag}
).
\label{eqn:GKLS}
\end{align}
The lowest-order term $K^{(0)}$ is the Zeno generator, given by
\begin{align}
&H^{(0)}=
\begin{pmatrix}
\omega & 0 & 0 & 0 & 0\\
0 & -\delta/2 & 0 & 0 & 0\\
0 & 0 & \delta/2 & 0 & 0\\
0 & 0 & 0 & 0 & 0\\
0 & 0 & 0 & 0 & 0
\end{pmatrix},
\nonumber\\
&\Gamma_i^{(0)}=\kappa,\quad
L_i^{(0)}=\ket{i}\bra{0}\quad
(i=1,2)
.
\end{align}
The first-order term $K^{(1)}$ provides an approximation usually discussed in terms of adiabatic elimination, which in the present case is given by 
\begin{align}
&H^{(1)}
=
\frac{1}{4}
\begin{pmatrix}
0 & 0 & 0 & 0 & 0\\
0 & -|g_1|^2 & -g_1^*g_2 & 0 & 0\\
0 & -g_1g_2^* & -|g_2|^2 & 0 & 0\\
0 & 0 & 0 & |g_1|^2+|g_2|^2 & 0\\
0 & 0 & 0 & 0 & 0
\end{pmatrix},
\nonumber
\displaybreak[0]
\\
&\Gamma_\pm^{(1)}=\pm\frac{1}{4}|g_1g_2|,\quad
L_\pm^{(1)}=\frac{
\rme^{-\rmi\phi_1}\ket{1}
\mp\rmi\rme^{-\rmi\phi_2}
\ket{2}
}{\sqrt{2}}
\bra{4},
\end{align}
where $g_{1,2}=|g_{1,2}|\rme^{\rmi\phi_{1,2}}$.
Notice here that these approximations are valid only for limited time ranges.
See Fig.~\ref{fig:DissipativeLambdaEvolution}.
The Zeno generator $K_\text{eff}^{(0)}=K^{(0)}$ is a good approximation only for times up to $t=O(\gamma)$, while the evolution with $K_\text{eff}^{(1)}=K^{(0)}+K^{(1)}/\gamma$ by adiabatic elimination starts to deviate from the true evolution for $t=O(\gamma^2)$.
The second- and third-order approximations $K^{(2)}$ and $K^{(3)}$ are given by
\begin{align}
&H^{(2)}
=
\frac{1}{8}
\delta
\begin{pmatrix}
0 & 0 & 0 & 0 & 0\\
0 & |g_1|^2 & 0 & 0 & 0\\
0 & 0 & -|g_2|^2 & 0 & 0\\
0 & 0 & 0 & -|g_1|^2+|g_2|^2 & 0\\
0 & 0 & 0 & 0 & 0
\end{pmatrix},
\nonumber
\displaybreak[0]
\\
&\Gamma_\pm^{(2)}
=\pm\frac{1}{4}\kappa(|g_1|^2+|g_2|^2),
\nonumber
\displaybreak[0]
\\
&
L_+^{(2)}=\ket{3}\bra{0},\quad
L_-^{(2)}=\frac{g_1^*\ket{1}
+
g_2^*\ket{2}
}{\sqrt{|g_1|^2+|g_2|^2}}
\bra{0},
\end{align}
and
\begin{align}
&H^{(3)}
=
\frac{1}{16}
(\delta^2-|g_1|^2-|g_2|^2)
\nonumber\\
&\qquad\quad
{}\times
\begin{pmatrix}
0 & 0 & 0 & 0 & 0\\
0 & -|g_1|^2 & -g_1^*g_2 & 0 & 0\\
0 & -g_1g_2^* & -|g_2|^2 & 0 & 0\\
0 & 0 & 0 & |g_1|^2+|g_2|^2 & 0\\
0 & 0 & 0 & 0 & 0
\end{pmatrix},
\nonumber
\displaybreak[0]
\\
&\Gamma_1^{(3)}
=+\frac{1}{4}\kappa\delta|g_1|^2,
\qquad\qquad\ \ \,\,%
L_1^{(3)}=\ket{1}\bra{0},
\nonumber
\displaybreak[0]
\\
&
\Gamma_2^{(3)}
=-\frac{1}{4}\kappa\delta|g_2|^2,
\qquad\qquad\ \ \,\,%
L_2^{(3)}=\ket{2}\bra{0},
\nonumber
\displaybreak[0]
\\
&\Gamma_3^{(3)}
=-\frac{1}{4}\kappa\delta(|g_1|^2-|g_2|^2),
\quad
L_3^{(3)}=\ket{3}\bra{0},
\nonumber
\displaybreak[0]
\\
&\Gamma_\pm^{(3)}
=\pm\frac{1}{16}|g_1g_2|(\delta^2-|g_1|^2-|g_2|^2),
\nonumber\\
&\qquad\qquad\quad\ %
L_\pm^{(3)}
=
\frac{
\rme^{-\rmi\phi_1}
\ket{1}
\mp\rmi\rme^{-\rmi\phi_2}
\ket{2}
}{\sqrt{2}}
\bra{4}.
\end{align}
These extend the valid time range up to $t= O(\gamma^3)$ and $t= O(\gamma^4)$, respectively.
In general, the $k$th-order adiabatic approximation $K_\text{eff}^{(k)}=\sum_{j=0}^kK^{(j)}/\gamma^j$ works well for times up to $t= O(\gamma^{k+1})$, and the nonperturbative adiabatic generator $K=K_\text{eff}^{(\infty)}$ works eternally, keeping the error $O(1/\gamma)$, as is clearly observed in Fig.~\ref{fig:DissipativeLambdaEvolution}.

For a nonvanishing $\delta$, it is generally impossible to get an analytical expression for the nonperturbative adiabatic generator $K$, but it can be estimated numerically.
For instance, for $\omega=\delta=g_1=g_2=\kappa=\kappa_0=1$, and $\gamma=10$, we get $K=K_\text{eff}^{(\infty)}$ in the GKLS form 
\begin{align}
K
={}&{-\rmi}[H,{}\bullet{}]
\nonumber\\
&{}-\frac{1}{2}\sum_i
\Gamma_i(
L_i^\dag L_i{}\bullet{}
+
{}\bullet{}L_i^\dag L_i
-
2L_i{}\bullet{}L_i^\dag
),
\label{eqn:GKLS2}
\end{align}
with
\begin{align}
H
&=
\begin{pmatrix}
1&0&0&0&0\\
0&-0.524&-0.025&0&0\\
0&-0.025&\hphantom{-}0.474&0&0\\
0&0&0&0.050&0\\
0&0&0&0&0
\end{pmatrix},
\nonumber
\displaybreak[0]
\\
\Gamma_1
&=\hphantom{{+}}1.000,
\quad
L_1
=\Bigl(\cos\theta\,\ket{1}-\rme^{\rmi\phi}\sin\theta\,\ket{2}\Bigr)\,\bra{0},
\nonumber
\displaybreak[0]
\\
\Gamma_2
&=\hphantom{{+}}0.995,
\quad
L_2
=\Bigl(\rme^{-\rmi\phi}\sin\theta\,\ket{1}+\cos\theta\,\ket{2}\Bigr)\,\bra{0},
\nonumber
\displaybreak[0]
\\
\Gamma_3
&=\hphantom{{+}}0.005,
\quad%
L_3
=\ket{3}\bra{0},
\nonumber
\\
\Gamma_\pm
&=\pm0.025,
\quad
L_\pm
=
\frac{
\ket{1}\mp\rmi\ket{2}}{\sqrt{2}}
\bra{4},
\label{eqn:Kinf}
\end{align}
where $\tan\theta=0.909$, $\tan\phi=0.029$.
To get this nonperturbative generator $K$ numerically, we used the adiabatic Bloch equation~(\ref{eqn:AdiabaticBloch}) as
\begin{equation}
\Omega_\ell=CP_\ell+S_\ell\Omega_\ell N_\ell-\frac{1}{\gamma}CS_\ell\Omega_\ell+\frac{1}{\gamma}S_\ell\Omega_\ell^2\equiv f(\Omega_\ell),
\end{equation}
and performed naive iterations over the function $f$, which for $\gamma=10$ converged quickly with the initial guess $\Omega_\ell^{(0)}=\langle C\rangle P_\ell=\sum_{n=0}^{n_\ell-1}S_\ell^n CN_\ell^nP_\ell$, which is the zeroth-order solution of $\Omega_\ell$ (there is no nilpotent $N_\ell$ in the present model and the initial guess we used was simply $CP_\ell$).
A more sophisticated algorithm with advanced convergence speed and guaranteed solution using Newton iteration is provided in Ref.~\cite{Lancaster}.
See Appendix~\ref{bounds} for the conditions for the existence and the uniqueness of the solution to the adiabatic Bloch equation~(\ref{eqn:AdiabaticBloch}) based on the Newton-Kantorovich theorem for the Newton iteration~\cite{Ortega}.
After obtaining $D_\ell=P_\ell\Omega_\ell P_\ell$ from $\Omega_\ell$, we also solved the conjugate adiabatic Bloch equation~(\ref{eqn:AdiabaticBlochRev}) numerically, constructed $U_\ell$ and $\tilde{U}_\ell$ through~(\ref{eqn:Uell}) and~(\ref{eqn:UellRev}), respectively, and applied the similarity transformation $(\tilde{U}_\ell U_\ell)^{1/2}$ to get $K_\ell$ from $D_\ell$ according to~(\ref{eqn:SWD}).
We can also solve the Bloch equations~(\ref{eqn:Bloch}) and~(\ref{eqn:BlochRev}) in the same way to obtain $U_\ell$ and $\tilde{U}_\ell$ directly, instead of solving~(\ref{eqn:AdiabaticBloch}) and~(\ref{eqn:AdiabaticBlochRev}) for $\Omega_\ell$ and $\tilde{\Omega}_\ell$.
Then, we can construct $K_\ell$ according to~(\ref{eqn:SWD}).

\begin{table}[b]
\caption{The spectra of $B$ and $\gamma B+C$ of the dissipative $\Lambda$ system~(\ref{eqn:DissipativeLambdaBC})--(\ref{eqn:DissipativeLambdaHL}) with $\delta=0$. Here, $g=\sqrt{|g_1|^2+|g_2|^2}$.}
\label{tab:SpectraEx1}
\begin{ruledtabular}
\begin{tabular}{cc}
\smallskip
$B$&$\gamma B+C$\\
\hline
&$0$\ \ (threefold degenerated)\vphantom{$\biggr)$}\\
&$\pm\frac{\rmi}{2}\,\Bigl(\sqrt{\gamma^2+g^2}-\gamma\Bigr)$\vphantom{$\biggr)$}\\
$0$&$-\kappa\pm\rmi\omega$\vphantom{$\biggr)$}\\
&$-\kappa
\pm\rmi\left[
\omega+\frac{1}{2}\,\Bigl(\sqrt{\gamma^2+g^2}-\gamma\Bigr)
\right]$\vphantom{$\biggr)$}\\
&$-2\kappa$\vphantom{$\biggr)$}\\
\hline
&${\pm\frac{\rmi}{2}}\,\Bigl(\gamma+\sqrt{\gamma^2+g^2}\Bigr)$\vphantom{$\biggr)$}\\
$\pm\rmi$&$\pm\rmi\sqrt{\gamma^2+g^2}$\vphantom{$\biggr)$}\\
&$-\kappa\pm\rmi\left[
\frac{1}{2}\,\Bigl(\gamma+\sqrt{\gamma^2+g^2}\Bigr)-\omega
\right]$\vphantom{$\biggr)$}\\
\hline
$-\frac{1}{2}\kappa_0\pm\rmi$&
$-\frac{1}{2}\gamma\kappa_0
\pm\frac{\rmi}{2}\,\Bigl(3\gamma-\sqrt{\gamma^2+g^2}\Bigr)$\vphantom{$\biggr)$}\\
\hline
&$-\frac{1}{2}\gamma\kappa_0\pm2\rmi\gamma$\vphantom{$\biggr)$}\\
$-\frac{1}{2}\kappa_0\pm2\rmi$&$-\frac{1}{2}\gamma\kappa_0
\pm\frac{\rmi}{2}\,\Bigl(3\gamma+\sqrt{\gamma^2+g^2}\Bigr)$\vphantom{$\biggr)$}\\
&$-\frac{1}{2}\gamma\kappa_0-\kappa\pm\rmi(2\gamma-\omega)$\vphantom{$\biggr)$}\\
\hline
$-\kappa_0$&
$-\gamma\kappa_0$\vphantom{$\biggr)$}
\end{tabular}
\end{ruledtabular}
\end{table}

One might have noticed that the perturbative terms presented above are all HP and TP, but not CP, except for the Zeno generator $K^{(0)}$, because of the non-positive-semidefinite Kossakowski matrices in the dissipators.
In the nonperturbative adiabatic generator $K$ in~(\ref{eqn:Kinf}), summing up all the perturbative contributions, there remains one negative eigenvalue $\Gamma_-=-0.025$ in the Kossakowski matrix.
It is associated with the strong decay from $\ket{4}$ to the $\Lambda$ subspace.
This negativity is not canceled by the dissipative part of the strong generator $\gamma B$: the total adiabatic generator $\gamma B+K$ has a negative eigenvalue $\tilde{\Gamma}_-=-6.22\times10^{-5}$ in its Kossakowski matrix with a Lindblad operator $\tilde{L}_-=(\cos\tilde{\theta}\,\ket{1}+\rmi\sin\tilde{\theta}\,\ket{1})\bra{4}$, where $\tan\tilde{\theta}=0.0025$.

If one computes $D$ for the present model, it is not CP even in the absence of the decays (i.e.~even for $\kappa_0=\kappa=0$).
It is turned into $K$ by the Schrieffer-Wolff transformation and becomes skew-Hermitian and CP\@.
The Schrieffer-Wolff transformation, however, does not amend CP in the presence of the decays.
The unitary part, on the other hand, is properly amended by the Schrieffer-Wolff transformation, even in the presence of the decays.
The decaying components anyway decay out, and the adiabatic evolution within the decoherence-free subspaces $\{\ket{1},\ket{2}\}$ and $\{\ket{3}\}$ are described by the Hamiltonian part $H$ of the resummed perturbative series in the long terms.
In any case, the error remains $O(1/\gamma)$ eternally, and the adiabatic evolution is well approximated by the effective generator $K$.

For $\delta=0$, analytical expressions are available.
The spectrum of $\gamma B+C$ is listed in Table~\ref{tab:SpectraEx1}, and the nonperturbative adiabatic generator $K$ is given in the GKLS form~(\ref{eqn:GKLS2}) with
\begin{widetext}
\begin{align}
H
&=
\omega\ket{0}\bra{0}
+\frac{1}{2}
\,\Bigl(
\sqrt{\gamma^2+g^2}-\gamma
\Bigr)
\begin{pmatrix}
0&0&0&0&0\\
0&-|g_1|^2/g^2&-g_1^*g_2/g^2&0&0\\
0&-g_1g_2^*/g^2&-|g_2|^2/g^2&0&0\\
0&0&0&1&0\\
0&0&0&0&0
\end{pmatrix},
\nonumber
\displaybreak[0]
\\
\Gamma_1
&=\kappa,
\qquad\qquad\qquad\qquad\qquad\qquad\quad\,%
L_1
=\frac{1}{g}\,\Bigl(
g_2\ket{1}
-
g_1\ket{2}
\Bigr)\,
\bra{0},
\nonumber
\\
\Gamma_2
&=\kappa\frac{\gamma^2+\gamma\sqrt{\gamma^2+g^2}+g^2+8\kappa^2}{2(\gamma^2+g^2+4\kappa^2)},
\quad
L_2
=\frac{1}{g}\,\Bigl(
g_1^*\ket{1}
+
g_2^*\ket{2}
\Bigr)\,\bra{0},
\nonumber
\\
\Gamma_3
&=\kappa\frac{\gamma^2-\gamma\sqrt{\gamma^2+g^2}+g^2}{2(\gamma^2+g^2+4\kappa^2)},
\qquad\quad\ \ \,%
L_3
=\ket{3}\bra{0},
\nonumber
\displaybreak[0]
\\
\Gamma_\pm
&=\pm\frac{1}{2}\,\Bigl(\sqrt{\gamma^2+g^2}-\gamma\Bigr)\,\frac{|g_1g_2|}{g^2},
\qquad\,%
L_\pm
=
\frac{1}{\sqrt{2}}
\,\Bigl(
\rme^{-\rmi\phi_1}\ket{1}\mp\rmi \rme^{-\rmi\phi_2}\ket{2}
\Bigr)\,\bra{4},
\label{eqn:Kana}
\end{align}
\end{widetext}
where $g=\sqrt{|g_1|^2+|g_2|^2}$.
Combined with the strong generator $\gamma B$, the Kossakowski matrix of the total adiabatic generator $\gamma B+K$ has the same spectrum $\{\Gamma_i\}$ as~(\ref{eqn:Kana}) except for the last two terms with $\Gamma_\pm$ and $L_\pm$, which are replaced by 
\begin{align}
\tilde{\Gamma}_\pm
={}&
\frac{1}{2}
\gamma\kappa_0
\left(
1
\pm
\sqrt{
1+
4\tan^2\phi\,\frac{|g_1g_2|^2}{g^4}
}
\right),
\nonumber\\
&
\tilde{L}_+
=
\Bigl(
c_1\rme^{-\rmi\phi_1}\ket{1}
-c_2\rme^{-\rmi\phi_2}\ket{2}
\Bigr)\,
\bra{4},
\nonumber\\
&
\tilde{L}_-
=
\Bigl(
c_2^*\rme^{-\rmi\phi_1}\ket{1}
+c_1^*\rme^{-\rmi\phi_2}\ket{2}
\Bigr)\,
\bra{4},
\label{eqn:Ex1gB+Kana}
\end{align}
where
\begin{gather}
\begin{cases}
\medskip
\displaystyle
c_1=
(
u_+
|g_2|
-
u_-
\rme^{\rmi\phi}
|g_1|
)/g
,\\
\displaystyle
c_2=
(
u_+
|g_1|
+
u_-\rme^{\rmi\phi}
|g_2|
)/g
,
\end{cases}
\tan\phi=\frac{\sqrt{\gamma^2+g^2}-\gamma}{2\gamma\kappa_0},
\nonumber
\displaybreak[0]
\\
u_\pm
=\sqrt{\frac{1}{2}\left(
1\pm\frac{1}{\sqrt{1+4\tan^2\phi\,|g_1g_2|^2/g^4}}
\right)
}.
\label{eqn:KhatPars}
\end{gather}
The eigenvalue $\tilde{\Gamma}_-$ is strictly negative, which is
\begin{equation}
\tilde{\Gamma}_-
=
-
\frac{|g_1g_2|^2}{16\gamma^3\kappa_0}
+ O(1/\gamma^5)
\label{eqn:GammaNegative}
\end{equation}
for large $\gamma$.

\subsection{Single-Qubit with Nilpotent}
\label{sec:ExampleNilpotent}
We can apply our formalism to open systems, even for a generator $B$ that admits a nilpotent.
Let us look at a simple qubit example,
\begin{gather}
B=-\frac{\rmi}{2}[X,{}\bullet{}]-(1-Z{}\bullet{}Z),
\\
C=-\rmi[X+Y,{}\bullet{}],
\end{gather}
where $X$, $Y$, and $Z$ are Pauli operators.
In a matrix representation, the generator $B$ is put in the Jordan normal form 
\begin{equation}
B
=
R
\begin{pmatrix}
-2&&&\\
&-1&\hphantom{-}1&\\
&\hphantom{-}0&-1&\\
&&&0
\end{pmatrix}
R^{-1},
\end{equation}
via a similarity transformation $R$.
The eigenvalue $-1$ is degenerate and accompanies a nilpotent in its eigenspace.
In this basis, the weak part $C$ of the generator is represented by
\begin{equation}
C
=
R
\begin{pmatrix}
0&-2&0&0\\
2&-2&2&0\\
2&-4&2&0\\
0&\hphantom{-}0&0&0
\end{pmatrix}
R^{-1}.
\end{equation}

This simple model is tractable analytically.
For instance, the spectrum of $\gamma B+C$ reads
\begin{equation}
\{
0,
-\gamma\pm2\rmi\sqrt{\gamma+2},
-2\gamma
\}.
\label{eqn:SpectrumQubit}
\end{equation}
Moreover, we can solve the adiabatic Bloch equation and get the nonperturbative adiabatic generator
\begin{align}
K
&=
\Bigl(\sqrt{\gamma^2+4\gamma+8}-\gamma\Bigr)\,
R
\begin{pmatrix}
 0 & 0 & 0 & 0 \\
 0 & -1 & 1 & 0 \\
 0 & -2 & 1 & 0 \\
 0 & 0 & 0 & 0
\end{pmatrix}
R^{-1}
\nonumber\\
&=-\frac{\rmi}{2}\left(\sqrt{\gamma^2+4\gamma+8}-\gamma\right)[X,{}\bullet{}].
\end{align}
Note that even though $K$ is endowed with the same block structure as $B$ they do not commute, $[B,K]\neq0$.
Observe also that $K$ is physical, i.e.~HP, TP, and CP, in this example.
The adiabatic generator $\gamma B+K$ is similar to the original generator $\gamma B+C$ as
\begin{equation}
\gamma B+K=W^{-1}(\gamma B+C)W
\end{equation}
with
\begin{equation}
W=
R
\begin{pmatrix}
\smallskip
1 & -\frac{2}{\sqrt{\gamma^2+4\gamma+8}} & \frac{2}{\sqrt{\gamma^2+4\gamma+8}} & 0 \\
\smallskip
 0 & 1 & 0 & 0 \\
\smallskip
 -\frac{2}{\gamma+2} & 1-\frac{\gamma+2}{\sqrt{\gamma^2+4\gamma+8}} & \frac{\gamma+2}{\sqrt{\gamma^2+4\gamma+8}} & 0 \\
 0 & 0 & 0 & 1
\end{pmatrix}
R^{-1},
\end{equation}
and they share the same spectrum~(\ref{eqn:SpectrumQubit}).

\begin{table}[b]
\caption{The spectra of $B$ and $\gamma B+C$ for the three-level system~(\ref{eqn:CountEx2})--(\ref{eqn:CountEx2HF}).}
\label{tab:SpectraEx3}
\begin{ruledtabular}
\begin{tabular}{c@{}c}
\smallskip
$B$&$\gamma B+C$\\
\hline
\multirow{2}{*}{$0$}&$0$\ \ (twofold degenerated)\\
&$-2$\\
\hline
$\pm\frac{\rmi}{3}$&$-\frac{1}{2}\pm\frac{\rmi}{3}\gamma$$\vphantom{\Bigl(}$\\
\hline
$\pm\frac{2\rmi}{3}$&$-\frac{1}{2}\pm\frac{2\rmi}{3}\gamma$$\vphantom{\Bigl(}$\\
\hline
$\pm\rmi$&$-1\pm\rmi\sqrt{\gamma^2-1}$$\vphantom{\Bigl(}$
\end{tabular}
\end{ruledtabular}
\end{table}
\subsection{Impossibility of Physical Generator}
\label{sec:Impossibility}
In the previous qubit example, $K$ is physical (HP, TP, and CP), but it is just a lucky case.
Indeed, in the first example (dissipative $\Lambda$ system), the adiabatic generator $K$ is not of proper physical structure.
We are sure about HP and TP of $K$, as proved in Sec.~\ref{sec:Physicality}, but CP is not guaranteed in general.
One might think that CP can be amended via an additional small similarity transformation on $\gamma B+K$ keeping the block structure of $B$.
However, it is generally impossible, as we prove here.

We provide a counterexample,
\begin{equation}
B=-\rmi[H_0,{}\bullet{}],\quad
C=-(1-L_0{}\bullet{}L_0^\dag),
\label{eqn:CountEx2}
\end{equation}
with
\begin{equation}
H_0=\frac{1}{3}
\begin{pmatrix}
0&0&0\\
0&1&0\\
0&0&3
\end{pmatrix},\quad
L_0=
\begin{pmatrix}
0&0&1\\
0&0&0\\
1&0&0
\end{pmatrix}.
\label{eqn:CountEx2HF}
\end{equation}
The strong generator $B$ has 7 spectral blocks,
\begin{equation}
B
=
R
\begin{pmatrix}
0&&&&&&&&\\
&0&&&&&&&\\
&&0&&&&&&\\
&&&-\rmi/3&&&&&\\
&&&&\rmi/3&&&&\\
&&&&&-2\rmi/3&&&\\
&&&&&&2\rmi/3&&\\
&&&&&&&-\rmi&\\
&&&&&&&&\rmi
\end{pmatrix}
R^{-1}.
\label{eqn:Ex3Block}
\end{equation}
All the sectors are nondecaying.
The spectrum of the total generator $\gamma B+C$ is given in Table~\ref{tab:SpectraEx3},
and decays are induced by the perturbation $C$ in the nondecaying eigenspaces of $B$.
For this model, the adiabatic generator $K$ is obtained via the generalized Schrieffer-Wolff transformation in the GKLS form~(\ref{eqn:GKLS2}) with
\begin{align}
&H=
\frac{1}{3}
\,\Bigl(
\gamma-\sqrt{\gamma^2-1}
\Bigr)
\begin{pmatrix}
1&0&0\\
0&0&0\\
0&0&-1
\end{pmatrix},
\nonumber
\displaybreak[0]
\\
&\Gamma_1=\Gamma_2=\frac{1}{2},
\quad
L_1=
\begin{pmatrix}
0&0&1\\
0&0&0\\
1&0&0
\end{pmatrix},
\quad
L_2=
\begin{pmatrix}
0&0&-\rmi\\
0&0&0\\
\rmi&0&0
\end{pmatrix},
\nonumber
\displaybreak[0]
\\
&\Gamma_\pm
=\pm\frac{1}{3\sqrt{3}}
\,\Bigl(
\gamma-\sqrt{\gamma^2-1}
\Bigr),
\nonumber\\
&\qquad\qquad\qquad\ \ \,%
L_\pm=
\begin{pmatrix}
\rme^{\pm\pi\rmi/3}&0&0\\
0&\rme^{\mp\pi\rmi/3}&0\\
0&0&-1
\end{pmatrix}.
\label{eqn:GKLSex3}
\end{align}
This $K$ is HP and TP, but not CP\@.

Now, we try to find an adiabatic generator $\tilde{K}$ that is endowed with the same block structure as $B$, shares the same spectrum with $\gamma B+C$, and is physical (HP, TP, and CP), via an additional similarity transformation on $\gamma B+K$.
Let us first impose HP, TP, and the block structure of $B$ on $\gamma B+\tilde{K}$.
Then, a possible adiabatic generator $\gamma B+\tilde{K}$ is constrained to
\begin{widetext}
\begin{equation}
\gamma B+\tilde{K}
=
R
\begin{pmatrix}
r_1&r_2&r_3&&&&&\\
r_4&r_5&r_6&&&&&\\
-r_1-r_4&-r_2-r_5&-r_3-r_6&&&&&&\\
&&&r_7+\rmi r_8&&&&&\\
&&&&r_7-\rmi r_8&&&&\\
&&&&&r_9+\rmi r_{10}&&&\\
&&&&&&r_9-\rmi r_{10}&&\\
&&&&&&&r_{11}+\rmi r_{12}&\\
&&&&&&&&r_{11}-\rmi r_{12}
\end{pmatrix}
R^{-1},
\end{equation}
\end{widetext}
parametrized by 12 real parameters $(r_1,\ldots,r_{12})$.
By further requiring that $\gamma B+\tilde{K}$ should have the same spectrum as $\gamma B+C$ listed in Table~\ref{tab:SpectraEx3}, we realize that the parameters should satisfy the conditions
\begin{gather}
r_7=r_9=-\frac{1}{2},\quad
r_{11}=-1,\\
r_8=-\frac{1}{3}\gamma,\quad
r_{10}=-\frac{2}{3}\gamma,\quad
r_{12}=-\sqrt{\gamma^2-1},
\end{gather}
and
\begin{gather}
(r_1+r_5)-(r_3+r_6)=-2,
\label{eqn:Cond1}
\displaybreak[0]
\\
(r_1-r_3)
(r_5-r_6)
-
(r_2-r_3)
(r_4-r_6)
=0.
\label{eqn:Cond2}
\end{gather}
The last two constraints are for the top-left $3\times3$ block to admit the eigenvalues $0$ and $-2$.
In this way, we are left with 4 free parameters.
By tuning the remaining 4 parameters, we try to make $\gamma B+\tilde{K}$ physical.
Since it is already required to be HP and TP, we try to achieve CP\@.
In terms of the remaining parameters, the spectrum of the Kossakowski matrix of $\gamma B+\tilde{K}$ is given by
\begin{widetext}
\begin{equation}
\Biggl\{
\frac{1}{2}r_2,
\frac{1}{2}r_3,
\frac{1}{2}r_4,
\frac{1}{2}r_6,
-\frac{1}{2}(r_1+r_4),
-\frac{1}{2}(r_2+r_5),
\pm\frac{1}{12}
\sqrt{
9(
r_1+r_5
+1
)^2
+
3(r_1-r_5+1)^2
+12\left(\gamma-\sqrt{\gamma^2-1}\right)^2
}
\Biggr\}.
\label{eqn:ChoiSpec}
\end{equation}
\end{widetext}
All these eigenvalues should be nonnegative for $\gamma B+\tilde{K}$ to be CP\@.
However, the last eigenvalue is strictly negative, and it is impossible to achieve the goal by tuning the parameters and to make $\gamma B+\tilde{K}$ physical.

This counterexample leads us to the following conclusion.
If we wish to find an adiabatic generator endowed with the physical structure (HP, TP, and CP), we have to sacrifice some of the axioms listed in the introduction.

\section{Conclusions}
\label{sec:Conclusions}
We have developed a general perturbation theory based on an iterated adiabatic theorem for arbitrary finite-dimensional quantum systems. Special cases previously known are given by Zeno dynamics, adiabatic elimination, Bloch generators, des Cloisaux generators, and by the Schrieffer-Wolff approach. Although we showed that an ideal effective generator cannot always be provided in open quantum systems, our generalization provides a good approach to highlight the eternal adiabatic resilience of quantum systems to perturbations. We were able to provide concise bounds for this.
Finally, we note that many of our theorems can be generalized easily to bounded operators on infinite-dimensional Hilbert spaces, provided that appropriate bounds on the spectral gap appearing in the reduced resolvent are assumed.

\begin{acknowledgments}
This research was funded in part by the Australian Research Council (project number FT190100106), and by the Top Global University Project from the Ministry of Education, Culture, Sports, Science and Technology (MEXT), Japan.
PF and SP were partially supported by Istituto Nazionale di Fisica Nucleare (INFN) through the project ``QUANTUM''. PF and SP acknowledge support by MIUR via PRIN 2017 (Progetto di Ricerca di Interesse Nazionale), project QUSHIP (2017SRNBRK). PF was partially supported by the Italian National Group of Mathematical Physics (GNFM-INdAM). PF and SP were partially supported by Regione Puglia and by QuantERA ERA-NET Cofund in Quantum Technologies (GA No.\ 731473), project PACE-IN\@.
HN is partly supported by the Institute for Advanced Theoretical and Experimental Physics, Waseda University and by Waseda University Grant for Special Research Projects (Project Number: 2020C-272).
KY was supported by the Grants-in-Aid for Scientific Research (C) (No.~18K03470) and for Fostering Joint International Research (B) (No.~18KK0073) both from the Japan Society for the Promotion of Science (JSPS).
\end{acknowledgments}

\appendix
\section{Key Formula for the Adiabatic Theorem}
\label{app:KeyFormula}
Here, we show the derivation of the key formula~(\ref{eqn:KeyRelation}) for the iterative application of the adiabatic theorem.
Recall first $(B-b_\ell)S_\ell=1-P_\ell$ in~(\ref{eqn:BSP}), satisfied by the reduced resolvent $S_\ell$ defined in~(\ref{eqn:Resolvent}).
Note also that
\begin{align}
&\rme^{(t-s)(\gamma B+C)}
(B-b_\ell)
\nonumber\\
&\ %
=-\frac{1}{\gamma}
\left(
\frac{\partial}{\partial s}
\bigl(
\rme^{(t-s)(\gamma B+C)}
\rme^{s(\gamma b_\ell +C)}
\bigr)
\right)
\rme^{-s(\gamma b_\ell +C)}.
\end{align}
Combining these relations, we have
\begin{align}
&
\rme^{(t-s)(\gamma B+C)}
(1-P_\ell)
\vphantom{\frac{1}{\gamma}}
\nonumber
\displaybreak[0]
\\
&\quad
=\rme^{(t-s)(\gamma B+C)}
(B-b_\ell )S_\ell 
\vphantom{\frac{1}{\gamma}}
\nonumber
\displaybreak[0]
\\
&\quad
=-\frac{1}{\gamma}
\left(
\frac{\partial}{\partial s}
\bigl(
\rme^{(t-s)(\gamma B+C)}
\rme^{s(\gamma b_\ell +C)}
\bigr)
\right)
\rme^{-s(\gamma b_\ell +C)}
S_\ell
\label{keystep}.
\end{align}
Then, for an arbitrary operator $A$, we get
\begin{align}
&\int_0^t\rmd s\,
\rme^{(t-s)(\gamma B+C)}
A P_\ell
\rme^{s(\gamma B+D_\ell)}
\nonumber
\\
&\quad
=
\int_0^t\rmd s\,
\rme^{(t-s)(\gamma B+C)}
P_\ell
A
P_\ell
\rme^{s(\gamma B+D_\ell )}
\nonumber
\\
&\qquad{}
+
\int_0^t\rmd s\,
\rme^{(t-s)(\gamma B+C)}
(1-P_\ell)
A
P_\ell
\rme^{s(\gamma B+D_\ell )}
\nonumber
\\
&\quad
=
\int_0^t\rmd s\,
\rme^{(t-s)(\gamma B+C)}
P_\ell
A
P_\ell
\rme^{s(\gamma B+D_\ell )}
\nonumber
\displaybreak[0]
\\
&\qquad{}
-\frac{1}{\gamma}
\int_0^t\rmd s
\left(
\frac{\partial}{\partial s}
\bigl(
\rme^{(t-s)(\gamma B+C)}
\rme^{s(\gamma b_\ell+C)}
\bigr)
\right)
\nonumber
\\[-1truemm]
&\qquad\qquad\qquad\qquad\qquad\quad{}\times
\rme^{-s(\gamma b_\ell+C)}
S_\ell 
A
P_\ell
\rme^{s(\gamma B+D_\ell)}
\nonumber
\displaybreak[0]
\\
&\quad
=
\int_0^t\rmd s\,
\rme^{(t-s)(\gamma B+C)}
P_\ell
A
P_\ell
\rme^{s(\gamma B+D_\ell )}
\nonumber
\displaybreak[0]
\\
&\qquad
{}-\frac{1}{\gamma}
\,\biggl[
\rme^{(t-s)(\gamma B+C)}
S_\ell 
A
P_\ell
\rme^{s(\gamma B+D_\ell )}
\biggr]_{s=0}^{s=t}
\nonumber
\displaybreak[0]
\\
&\qquad{}
+\frac{1}{\gamma}
\int_0^t\rmd s\,
\rme^{(t-s)(\gamma B+C)}
\rme^{s(\gamma b_\ell+C)}
\nonumber\\[-2truemm]
&\qquad\qquad\qquad\qquad{}\times
\frac{\partial}{\partial s}
\bigl(
\rme^{-s(\gamma b_\ell+C)}
S_\ell 
A
P_\ell
\rme^{s(\gamma B+D_\ell)}
\bigr)
\nonumber
\displaybreak[0]
\\
&\quad
=
\int_0^t\rmd s\,
\rme^{(t-s)(\gamma B+C)}
P_\ell
A
P_\ell
\rme^{s(\gamma B+D_\ell )}
\nonumber
\displaybreak[0]
\\
&\qquad{}
+\frac{1}{\gamma}
\rme^{t(\gamma B+C)}
S_\ell
A
P_\ell
-\frac{1}{\gamma}
S_\ell
A
P_\ell
\rme^{t(\gamma B+D_\ell )}
\nonumber
\displaybreak[0]
\\
&\qquad{}
-\frac{1}{\gamma}
\int_0^t\rmd s\,
\rme^{(t-s)(\gamma B+C)}
\mathcal{K}_\ell(A)P_\ell
\rme^{s(\gamma B+D_\ell )},
\end{align}
where $\mathcal{K}_\ell$ is defined in~(\ref{eq:Kelldef}).
The key formula~(\ref{eqn:KeyRelation}) is thus obtained.

\section{Bounding the Last Term of~(\ref{key})}
\label{app:Bounding2.16}
We show that the last term of~(\ref{key}) decays as $n\to+\infty$.
To show this, let us bound $A_\ell^{(n)}/\gamma^n=\mathcal{K}_\ell^n(C-D_\ell)/\gamma^n$, where $\mathcal{K}$ is defined in~(\ref{eq:Kelldef}). 
Recall that there exists an integer $n_\ell\ge1$ such that $N_\ell^{n_\ell}=0$.
This limits the highest possible power of $\gamma$ in the expansion of $\mathcal{K}_\ell^n$ to $n-\lfloor n/n_\ell\rfloor$, where $\lfloor x\rfloor$ is the largest integer less than or equal to $x$.
This is because, in the expansion of $\mathcal{K}_\ell^n$, the nilpotent $N_\ell$ can repeat only $n_\ell-1$ times sequentially and $D_\ell$ should interrupt the sequence.
The highest-order terms look like $\gamma^{n-\lfloor n/n_\ell\rfloor}S_\ell^n{}\bullet{}N_\ell^pD_\ell(N_\ell^{n_\ell-1}D_\ell)^{\lfloor n/n_\ell\rfloor-1}N_\ell^q$ with integers $p$ and $q$ satisfying $p,q\le n_\ell-1$ and $p+q=n-(\lfloor n/n_\ell\rfloor-1)n_\ell-1$.
Therefore, $A_\ell^{(n)}$ is bounded by
\begin{widetext}
\begin{align}
\|A_\ell^{(n)}\|
&\le\sum_{r=0}^{n-\lfloor n/n_\ell\rfloor}
\begin{pmatrix}
n\\
r
\end{pmatrix}
(\|C\|\|S_\ell\|+\|S_\ell\|\|D_\ell\|)^{n-r}
(\gamma\|S_\ell\|\|N_\ell\|)^r\|C-D_\ell\|.
\intertext{It is a rough bound since it is overcounting also vanishing terms containing $N_\ell^m$ with $m>n_\ell-1$, but this suffices for our purpose. For $\gamma>1$, it is further bounded by}
&\le\gamma^{n-\lfloor n/n_\ell\rfloor}
\|S_\ell\|^n
\sum_{r=0}^{n-\lfloor n/n_\ell\rfloor}
\begin{pmatrix}
n\\
r
\end{pmatrix}
(\|C\|+\|D_\ell\|)^{n-r}
\|N_\ell\|^r\|C-D_\ell\|
\nonumber
\displaybreak[0]
\\
&\le\gamma^{n-\lfloor n/n_\ell\rfloor}
\,\Bigl(
\|S_\ell\|(\|C\|+\|D_\ell\|+\|N_\ell\|)
\Bigr)^n
\|C-D_\ell\|.
\intertext{Since $(n+1)/n_\ell-1\le\lfloor n/n_\ell\rfloor\le n/n_\ell$,}
&\le\gamma^{n-(n+1)/n_\ell+1}
\,\Bigl(
\|S_\ell\|(\|C\|+\|D_\ell\|+\|N_\ell\|)
\Bigr)^n
\|C-D_\ell\|
\nonumber
\displaybreak[0]
\\
&=\gamma^{n-1/n_\ell+1}
\left(
\frac{
[
\|S_\ell\|(\|C\|+\|D_\ell\|+\|N_\ell\|)
]^{n_\ell}
}{\gamma}
\right)^{n/n_\ell}
\|C-D_\ell\|.
\end{align}
\end{widetext}
Therefore, $\|A_\ell^{(n)}\|/\gamma^n\to0$ as $n\to+\infty$, provided $\gamma>\max\{1,[\|S_\ell\|(\|C\|+\|D_\ell\|+\|N_\ell\|)]^{n_\ell}\}$.

\section{Link with Bloch's Perturbation Theory}
\label{app:GeneralizedBlochEq}
We want to translate our adiabatic Bloch equation~(\ref{eqn:AdiabaticBloch})--(\ref{eqn:AB2}) for $\Omega_\ell$ into the equation for the similarity transformation $U_\ell$ defined in~(\ref{eqn:Uell}).
This will show that our theory  is equivalent to Bloch's perturbation theory in the unitary case~\cite{Bloch} and generalizes it to the nonunitary case.

Let us first try to invert the relation~(\ref{eqn:Uell}) between $U_\ell$ and $\Omega_\ell$, i.e.,
\begin{equation}
U_\ell
=P_\ell-\frac{1}{\gamma}S_\ell\Omega_\ell.
\label{eqn:UellApp}
\end{equation}
It yields $S_\ell\Omega_\ell/\gamma=P_\ell-U_\ell$.
We use it to replace $\Omega_\ell$ with $U_\ell$ in our adiabatic Bloch equation~(\ref{eqn:AdiabaticBloch}),
\begin{align}
\Omega_\ell
&=\frac{1}{\gamma}S_\ell\Omega_\ell^{2}
-\frac{1}{\gamma}CS_\ell\Omega_\ell
+S_\ell\Omega_\ell N_\ell
+CP_\ell
\nonumber
\displaybreak[0]
\\
&=(P_\ell-U_\ell)\Omega_\ell
-C(P_\ell-U_\ell)
+\gamma(P_\ell-U_\ell)N_\ell
+CP_\ell
\vphantom{\frac{1}{\gamma}}
\nonumber\displaybreak[0]\\
&=CU_\ell
+(P_\ell-U_\ell)
(
\Omega_\ell
+\gamma N_\ell
)
\vphantom{\frac{1}{\gamma}}
\nonumber\\
&=CU_\ell
-(1-P_\ell)
U_\ell(
\Omega_\ell
+\gamma N_\ell
),
\vphantom{\frac{1}{\gamma}}
\label{eqn:U2OmegaA}
\end{align}
where we have used $P_\ell U_\ell=P_\ell$ from~(\ref{eqn:UPPU}).
This implies
\begin{equation}
P_\ell\Omega_\ell
=P_\ell CU_\ell.
\label{eqn:POmega2PCU}
\end{equation}
Therefore, by inserting it back into the right-hand side of~(\ref{eqn:U2OmegaA}) and by noting $U_\ell P_\ell=U_\ell$ from~(\ref{eqn:UPPU}), we get
\begin{equation}
\Omega_\ell
=CU_\ell
-(1-P_\ell)U_\ell(
CU_\ell
+\gamma N_\ell
).
\label{eqn:U2Omega}
\end{equation}
This is the inversion of the relation~(\ref{eqn:UellApp}).

By inserting this expression into the right-hand side of the relation~(\ref{eqn:UellApp}), we obtain the equation for $U_\ell$ as
\begin{equation}
U_\ell=P_\ell-\frac{1}{\gamma}S_\ell(CU_\ell-U_\ell CU_\ell)
+S_\ell U_\ell N_\ell,
\label{eqn:BlochApp}
\end{equation}
with
\begin{equation}
U_\ell P_\ell=U_\ell.
\label{eqn:UPApp}
\end{equation}
These equations are presented in~(\ref{eqn:Bloch}) and~(\ref{eqn:UP}) of the main text.
Note that  equation~(\ref{eqn:BlochApp}) automatically reproduces one of the two properties of $U_\ell$ in~(\ref{eqn:UPPU}), $P_\ell U_\ell=P_\ell$, while the other one $U_\ell P_\ell=U_\ell$ is independent of~(\ref{eqn:BlochApp}).
We need~(\ref{eqn:UPApp}) in addition to the equation~(\ref{eqn:BlochApp}) to characterize $U_\ell$.

When $B$ and $C$ are Hamiltonians (multiplied by $-\rmi$), there is no nilpotent $N_\ell$ in $B$, and the equation~(\ref{eqn:BlochApp}) for $U_\ell$ is nothing but the well-known Bloch equation~\cite{Bloch}.
Our equation~(\ref{eqn:BlochApp}) generalizes Bloch's equation to the case where $B$ and $C$ are not skew-Hermitian  and $B$ might be even nondiagonalizable.
In particular, our formalism can describe noisy quantum dynamics.

Let us check the validity of the results just obtained.
First, we assume that $\Omega_\ell$ satisfies our adiabatic Bloch equation~(\ref{eqn:AdiabaticBloch}) with~(\ref{eqn:AB2}) and show that $U_\ell$ introduced through the relation~(\ref{eqn:UellApp}) solves the generalized Bloch equation~(\ref{eqn:BlochApp}).
Before starting to show it, note that our adiabatic Bloch equation~(\ref{eqn:AdiabaticBloch}) multiplied by $P_\ell$ from the left yields
\begin{equation}
-P_\ell\left(1+\frac{1}{\gamma}CS_\ell\right)\Omega_\ell
+P_\ell CP_\ell
=0.
\label{eqn:PCP}
\end{equation}
Now, by inserting the relation~(\ref{eqn:UellApp}) for $U_\ell$,
\begin{align}
&
U_\ell-P_\ell+\frac{1}{\gamma}S_\ell(CU_\ell-U_\ell CU_\ell)
-S_\ell U_\ell N_\ell
\nonumber
\\
&\ %
=
\left(P_\ell-\frac{1}{\gamma}S_\ell\Omega_\ell\right)
-P_\ell
\nonumber
\displaybreak[0]
\\
&\qquad{}
+\frac{1}{\gamma}S_\ell\,\biggl[
C\left(P_\ell-\frac{1}{\gamma}S_\ell\Omega_\ell\right)
\nonumber
\displaybreak[0]
\\
&\qquad\qquad\quad\ \,{}
-
\left(P_\ell-\frac{1}{\gamma}S_\ell\Omega_\ell\right)
C\left(P_\ell-\frac{1}{\gamma}S_\ell\Omega_\ell\right)
\biggr]
\displaybreak[0]
\nonumber\\
&\qquad\qquad\qquad\qquad\qquad\qquad{}
-S_\ell
\left(P_\ell-\frac{1}{\gamma}S_\ell\Omega_\ell\right)
N_\ell
\displaybreak[0]
\nonumber\\
&\ %
=
-\frac{1}{\gamma}S_\ell\,\biggl[
\Omega_\ell
-
C\left(
P_\ell
-
\frac{1}{\gamma}S_\ell\Omega_\ell
\right)
\nonumber\\
&\qquad\qquad\quad{}
-
\frac{1}{\gamma}S_\ell\Omega_\ell
\left(
P_\ell
C
P_\ell
-
\frac{1}{\gamma}
P_\ell
C
S_\ell\Omega_\ell
\right)
-S_\ell\Omega_\ell
N_\ell
\biggr]
\nonumber\displaybreak[0]\\
&\ %
=
-\frac{1}{\gamma}S_\ell\,\biggl[
\Omega_\ell
-
C\left(
P_\ell
-
\frac{1}{\gamma}S_\ell\Omega_\ell
\right)
\nonumber\\
&\qquad\qquad\qquad\qquad\qquad\qquad\quad{}
-
\frac{1}{\gamma}S_\ell\Omega_\ell
P_\ell\Omega_\ell
-S_\ell\Omega_\ell
N_\ell
\biggr]
\nonumber\displaybreak[0]\\
&\ %
=
\frac{1}{\gamma}S_\ell
\left[
\frac{1}{\gamma}S_\ell\Omega_\ell^2
-
\left(
1+\frac{1}{\gamma}CS_\ell
\right)
\Omega_\ell
+
CP_\ell
+S_\ell\Omega_\ell N_\ell
\right]
\nonumber
\displaybreak[0]
\\
&\ %
=0.
\vphantom{\frac{1}{\gamma}}
\end{align}
We have used $S_\ell P_\ell=0$ and $\Omega_\ell=\Omega_\ell P_\ell$ from~(\ref{eqn:AB2}) for the second equality, used~(\ref{eqn:PCP}) to get the third equality, and used our adiabatic Bloch equation~(\ref{eqn:AdiabaticBloch}) for the last equality.
This proves that the generalized Bloch equation~(\ref{eqn:BlochApp}) is satisfied.
Equation~(\ref{eqn:UPApp}) also follows from the definition of $U_\ell$ in~(\ref{eqn:UellApp}) and $\Omega_\ell P_\ell=\Omega_\ell$ from~(\ref{eqn:AB2}).

The converse is also true.
We now assume that $U_\ell$ satisfies the generalized Bloch equation~(\ref{eqn:BlochApp}) with~(\ref{eqn:UPApp}) and show that $\Omega_\ell$ introduced through the relation~(\ref{eqn:U2Omega}) solves our Bloch equation~(\ref{eqn:AdiabaticBloch}).
By inserting the relation~(\ref{eqn:U2Omega}) for $\Omega_\ell$,
\begin{align}
&\frac{1}{\gamma}S_\ell\Omega_\ell^{2}-\left(1+\frac{1}{\gamma}CS_\ell\right)\Omega_\ell+CP_\ell
+S_\ell\Omega_\ell N_\ell
\nonumber
\\
&\ %
=\frac{1}{\gamma}S_\ell
[
CU_\ell
-(1-P_\ell)U_\ell(
CU_\ell
+\gamma N_\ell
)
]^2
\nonumber
\\
&\ \ \quad
{}
-\left(1+\frac{1}{\gamma}CS_\ell\right)
[
CU_\ell
-(1-P_\ell)U_\ell(
CU_\ell
+\gamma N_\ell
)
]
\nonumber
\\
&\ \ \quad
{}
+CP_\ell
+S_\ell
[
CU_\ell
-(1-P_\ell)U_\ell(
CU_\ell
+\gamma N_\ell
)
]
N_\ell
\vphantom{\frac{1}{\gamma}}
\nonumber
\displaybreak[0]
\\
&\ %
=
\frac{1}{\gamma}
S_\ell
[
CU_\ell
-U_\ell(
CU_\ell
+\gamma N_\ell
)
]
CU_\ell
\nonumber
\displaybreak[0]
\\
&\ \ \quad{}
-
CU_\ell
+(1-P_\ell)U_\ell(
CU_\ell
+\gamma N_\ell
)
\vphantom{\frac{1}{\gamma}}
\nonumber
\displaybreak[0]
\\
&\ \ \quad{}
-
C
\frac{1}{\gamma}
S_\ell
[
CU_\ell
-U_\ell(
CU_\ell
+\gamma N_\ell
)
]
\nonumber
\displaybreak[0]
\\
&\ \ \quad{}
+CP_\ell
+
\frac{1}{\gamma}S_\ell
[
CU_\ell
-U_\ell(
CU_\ell
+\gamma N_\ell
)
]
\gamma N_\ell
\vphantom{\frac{1}{\gamma}}
\nonumber
\displaybreak[0]
\\
&\ %
=
(P_\ell-U_\ell)
CU_\ell
-
CU_\ell
+(1-P_\ell)U_\ell(
CU_\ell
+\gamma N_\ell
)
\vphantom{\frac{1}{\gamma}}
\nonumber\\
&\ \ \quad{}
-
C
(P_\ell-U_\ell)
+CP_\ell
+\gamma
(P_\ell-U_\ell)
N_\ell
\nonumber\displaybreak[0]\\
&\ %
=0.
\vphantom{\frac{1}{\gamma}}
\end{align}
We have used $S_\ell(1-P_\ell)=S_\ell$ and $U_\ell(1-P_\ell)=0$ from~(\ref{eqn:UPApp}) for the second equality, used the generalized Bloch equation~(\ref{eqn:BlochApp}) to get the third equality, used $P_\ell U_\ell=P_\ell$, which follows from the generalized Bloch equation~(\ref{eqn:BlochApp}), for the last equality.
This proves that our adiabatic Bloch equation~(\ref{eqn:AdiabaticBloch}) is satisfied.
Equation~(\ref{eqn:AB2}) also follows from the relation~(\ref{eqn:U2Omega}) and $U_\ell P_\ell=U_\ell$ from~(\ref{eqn:UPApp}).

Finally, let us also check that~(\ref{eqn:UellApp}) and~(\ref{eqn:U2Omega})
are indeed the inverses of each other, provided that both Bloch equations~(\ref{eqn:AdiabaticBloch})--(\ref{eqn:AB2}) and~(\ref{eqn:BlochApp})--(\ref{eqn:UPApp}) hold:
by inserting~(\ref{eqn:U2Omega}) for $\Omega_\ell$ into the right-hand side of~(\ref{eqn:UellApp}) we immediately get
\begin{align}
P_\ell-\frac{1}{\gamma}S_\ell\Omega_\ell
&=
P_\ell-\frac{1}{\gamma}S_\ell
[
CU_\ell
-(1-P_\ell)U_\ell(
CU_\ell
+\gamma N_\ell
)
]
\nonumber\\
&=U_\ell,
\end{align}
thanks to the generalized Bloch equation~(\ref{eqn:BlochApp}), while by inserting~(\ref{eqn:UellApp}) for $U_\ell$ into the right-hand side of~(\ref{eqn:U2Omega}) we get
\begin{align}
&CU_\ell
-(1-P_\ell)U_\ell(
CU_\ell
+\gamma N_\ell
)
\nonumber\displaybreak[0]\\
&%
=
C
\left(
P_\ell-\frac{1}{\gamma}S_\ell\Omega_\ell
\right)
\nonumber\displaybreak[0]\\
&\quad{}
-(1-P_\ell)\left(
P_\ell-\frac{1}{\gamma}S_\ell\Omega_\ell
\right)
\left[
C
\left(
P_\ell-\frac{1}{\gamma}S_\ell\Omega_\ell
\right)
+\gamma N_\ell
\right]
\nonumber
\displaybreak[0]
\\
&%
=
C
\left(
P_\ell-\frac{1}{\gamma}S_\ell\Omega_\ell
\right)
\nonumber
\displaybreak[0]
\\
&\quad{}
+\frac{1}{\gamma}S_\ell\Omega_\ell
\left[
P_\ell
C
\left(
P_\ell-\frac{1}{\gamma}S_\ell\Omega_\ell
\right)
+\gamma N_\ell
\right]
\nonumber
\\
&%
=
C
\left(
P_\ell-\frac{1}{\gamma}S_\ell\Omega_\ell
\right)
+\frac{1}{\gamma}S_\ell\Omega_\ell
(
P_\ell
\Omega_\ell
+\gamma N_\ell
)
\nonumber
\displaybreak[0]
\\
&%
=
\frac{1}{\gamma}S_\ell\Omega_\ell^2
-\frac{1}{\gamma}CS_\ell\Omega_\ell
+CP_\ell
+S_\ell\Omega_\ell N_\ell
\nonumber\displaybreak[0]\\
&%
=
\Omega_\ell,
\vphantom{\frac{1}{\gamma}}
\end{align}
where we have used~(\ref{eqn:PCP}), which follows from our Bloch equation~(\ref{eqn:AdiabaticBloch}).
Everything is thus consistent.

\section{Solvability of the Adiabatic Bloch Equations}
\label{bounds}
For a given $\ell$, the adiabatic Bloch equations~(\ref{eqn:AdiabaticBloch}) and~(\ref{eqn:Bloch}) for $\Omega_\ell$ and $U_\ell$, respectively, are quadratic matrix equations.
Lancaster and Rokne~\cite{Lancaster} studied the existence and the uniqueness problem of a similar quadratic equation using the Newton-Kantorovich theorem~\cite{Ortega}.
We can follow similar proofs for the adiabatic Bloch equations~(\ref{eqn:AdiabaticBloch}) and~(\ref{eqn:Bloch}) using Ref.~\cite{Ortega} directly.
It shows the existence of a solution constructively by a converging Newton iteration finding a solution of the equation.
Let us show here the solvability of the adiabatic Bloch equation~(\ref{eqn:Bloch}) for the wave operator $U_\ell$.
We can also analyze the other adiabatic Bloch equation~(\ref{eqn:AdiabaticBloch}) for $\Omega_\ell$ in the same way.
Strictly speaking the adiabatic Bloch equation is a set of coupled equations~(\ref{eqn:Bloch})--(\ref{eqn:UP}).
We will see that the Newton iteration preserves the latter condition~(\ref{eqn:UP}), so we can solve both equations simultaneously.

The adiabatic Bloch equation~(\ref{eqn:Bloch}) for the wave operator $U_\ell$ is a quadratic matrix equation in $X=U_\ell$ of the form
\begin{equation}
\mathcal{F}(X)
=X
-S_\ell X N_\ell
+\frac{1}{\gamma}S_\ell(CX-XCX)
-P_\ell=0.
\end{equation}
The (Fr\'echet) derivative of $\mathcal{F}(X)$ reads
\begin{align}
\mathcal{F}'_X(A)
=
A-S_\ell AN_\ell
+\frac{1}{\gamma}S_\ell(CA-XCA-ACX).
\end{align}
The derivative $\mathcal{F}_X'$ is invertible for large $\gamma$,
\begin{equation}
(\mathcal{F}_X')^{-1}
=
\left(\mathcal{I}+\frac{1}{\gamma}\mathcal{G}_X\right)^{-1}
=\mathcal{I}^{-1}\left(1+\frac{1}{\gamma}\mathcal{G}_X\mathcal{I}^{-1}\right)^{-1},
\label{eqn:F'inv}
\end{equation}
where
\begin{gather}
\mathcal{I}(A)=A-S_\ell AN_\ell,
\quad
\mathcal{I}^{-1}(A)=\sum_{n=0}^{n_\ell-1}S_\ell^nAN_\ell^n,
\\
\mathcal{G}_X(A)
=S_\ell(CA-XCA-ACX).
\end{gather}
The Newton iteration is then given by
\begin{equation}
X_{k+1}=X_k-(\mathcal{F}'_{X_k})^{-1}(\mathcal{F}(X_k)).	
\label{eqn:NewtonIteration}
\end{equation}
It is reasonable to choose the zeroth-order solution of the perturbative equation as an initial guess.
With
\begin{equation}
X_0=U_\ell^{(0)}=\mathcal{I}^{-1}(P_\ell)=P_\ell,
\end{equation}
we have
\begin{equation}
\mathcal{F}(X_0)=\frac{1}{\gamma}S_\ell C P_\ell, 
\end{equation}
and
\begin{equation}
\mathcal{G}(X_0)(A)=S_\ell(CA-ACP_\ell).
\end{equation}

Explicit bounds are readily obtained from geometric series:
\begin{gather}
\|\mathcal{I}^{-1}\|
\le
\sum_{n=0}^{n_\ell-1}
(\|S_\ell\|\|N_\ell\|)^n
=\frac{1-(\|S_\ell\|\|N_\ell\|)^{n_\ell}}{1-\|S_\ell\|\|N_\ell\|}
\equiv\mu_\ell,
\\
\|\mathcal{F}(X_0)\| \leq \frac{1}{\gamma} \|S_\ell\| \|C\| \|P_\ell\|, 
\\
\|\mathcal{G}_{X_0}\| \leq 2 \|S_\ell\|\|C\|\|P_\ell\|,
\end{gather}
where we have used $\|P_\ell\|\geq 1$. Therefore,
\begin{align}
\|(\mathcal{F}_{X_0}')^{-1}\|
&\le
\frac{\|\mathcal{I}^{-1}\|}{
1-\frac{1}{\gamma}\|\mathcal{G}_{X_0}\|\|\mathcal{I}^{-1}\|
}
\nonumber
\displaybreak[0]
\\
&\le
\frac{\mu_\ell}{
1-\frac{2}{\gamma}\mu_\ell\|S_\ell\|\|C\|\|P_\ell\|
}
\equiv\beta_\ell,
\end{align}
\begin{equation}
\|(\mathcal{F}_{X_0}')^{-1}(\mathcal{F}(X_0))\|
\le
\frac{1}{\gamma}
\frac{
\mu_\ell
\|S_\ell\|\|C\|\|P_\ell\|
}{
1-\frac{2}{\gamma}\mu_\ell\|S_\ell\|\|C\|\|P_\ell\|
}\equiv\nu_\ell.
\end{equation}
Moreover, since
\begin{equation}
\mathcal{F}_{X}'(A)
-\mathcal{F}_{Y}'(A)
=
-\frac{1}{\gamma}S_\ell
[(X-Y)CA+AC(X-Y)],
\end{equation}
we have
\begin{equation}
\|\mathcal{F}'_X-\mathcal{F}'_Y\|
\le\frac{2}{\gamma }\|S_\ell\|\|C\|\|X-Y\|
\le L_\ell\|X-Y\|,
\end{equation}
with
\begin{equation}
L_\ell=\frac{2}{\gamma}\|S_\ell\|\|C\|\|P_\ell\|.
\end{equation}

According to Ref.~\cite{Ortega}, if 
\begin{equation}
h_\ell=\beta_\ell L_\ell\nu_\ell\le \frac{1}{2}, 
\end{equation}
there is a solution of $\mathcal{F}(X)=0$ within 
\begin{equation}
\|X-X_0\|\le\Theta_\ell=\frac{1-\sqrt{1-2h_\ell}}{\beta_\ell L_\ell}. 
\end{equation}
Moreover, there is at most  one solution within 
\begin{equation}
\|X-X_0\|<\Xi_\ell=\frac{1+\sqrt{1-2h_\ell}}{\beta_\ell L_\ell}.
\end{equation}
Finally, the convergence is at least quadratic if $h_\ell<1/2$.

In the present case, 
\begin{equation}
h_\ell=\beta_\ell L_\ell\nu_\ell
=
\frac{1}{\gamma^2}
\frac{
2\mu_\ell^2
\|S_\ell\|^2\|C\|^2\|P_\ell\|^2
}{
\left(1-\frac{2}{\gamma}\mu_\ell\|S_\ell\|\|C\|\|P_\ell\|\right)^2
},
\end{equation}
and
\begin{equation}
\Theta_\ell
=
\frac{
1-\sqrt{
1-\gamma_\ell/\gamma
}
}{
1+\sqrt{1-\gamma_\ell/\gamma}
}=\Xi_\ell^{-1},
\label{eqn:ThetaXi}
\end{equation}
with
\begin{equation}
\gamma_\ell= 4\mu_\ell\|S_\ell\|C\|\|P_\ell\|.
\end{equation}
The condition $h_\ell\le1/2$ for the solvability of the Bloch equation~(\ref{eqn:Bloch}) requires
\begin{equation}
\gamma\geq \gamma_\ell.
\label{eqn:x}
\end{equation}
Under this condition, a solution $U_\ell$ exists within 
\begin{equation}
\|U_\ell-P_\ell\|\le\Theta_\ell = O(1/\gamma),
\label{eqn:Ubound}
\end{equation}
and there is at most one solution within 
\begin{equation}
\|U_\ell-P_\ell\| < \Xi_\ell = O(\gamma). 
\end{equation}

We note that $X_0=X_0 P_\ell$. 
Furthermore, since $\mathcal{F}$ contains right-multiplication with only $N_\ell$, it preserves $X=XP_\ell$, i.e., $\mathcal{F}(X)=\mathcal{F}(X)P_\ell$.
The same holds for $F'_{X}(X)$ because it only contains right-multiplication by $N_\ell$ and $CX$, i.e., $\mathcal{F}'_X(X)=\mathcal{F}'(X)P_\ell$.
Therefore, the Newton iteration~(\ref{eqn:NewtonIteration}) preserves this property, and the limit $X_\infty$ fulfills both $\mathcal{F}(X_\infty)=0$ and $X_\infty=X_\infty P_\ell$. 
The solution $U_\ell=X_\infty$ obtained by the Newton iteration satisfies~(\ref{eqn:UP}).
In addition, the small distance $O(1/\gamma)$ from the initial guess $X_0=P_\ell$  justifies the perturbative approach taken in Sec.~\ref{Perturbative}\@.

Finally, the bound on $U_\ell$ in~(\ref{eqn:Ubound}) allows us to estimate the size of the adiabatic generator $D_\ell$.
Recalling that $D_\ell=P_\ell CU_\ell$, its norm is bounded by
\begin{align}
\|D_\ell\|
&=\|P_\ell CU_\ell\|
\nonumber\\
&\le\|P_\ell\|\|C\|\,\Bigl(1+\|U_\ell-P_\ell\|\Bigr)\,\|P_\ell\|
\nonumber\\
&\le\frac{2\|C\|\|P_\ell\|^2}{1+\sqrt{1-\gamma_\ell/\gamma}}.
\end{align}

\section{Eternal Bounds}
\label{app:EternalBounds}
We can also work on the conjugate Bloch equation~(\ref{eqn:BlochRev}) for $\tilde{U}_\ell$, and get
\begin{equation}
\|\tilde{U}_\ell-P_\ell\|
\le\Theta_\ell,
\label{eqn:UboundRev}
\end{equation}
with the same $\Theta_\ell$ given in~(\ref{eqn:ThetaXi}).
This and the bound on $U_\ell$ in~(\ref{eqn:Ubound}) allow us to explicitly bound the norm distance between the approximate adiabatic evolution $\rme^{t(\gamma B+K)}$ and the true evolution $\rme^{t(\gamma B+C)}$ eternally.

The similarity between the generators $\gamma B+C$ and $\gamma B+K$ in~(\ref{eqn:SimilarK}) implies the similarity between the evolutions $\rme^{t(\gamma B+C)}$ and $\rme^{t(\gamma B+K)}$.
The difference between the two evolutions is then estimated to be
\begin{align}
&\rme^{t(\gamma B+C)}
-
\rme^{t(\gamma B+K)}
\nonumber\\
&\quad
=
\rme^{t(\gamma B+C)}
-
W^{-1}\rme^{t(\gamma B+C)}W
\nonumber\\
&\quad
=
-\rme^{t(\gamma B+C)}(W-1)
-(W^{-1}-1)\rme^{t(\gamma B+C)}W
\nonumber\\
&\quad
=
-\sum_\ell
\rme^{t(\gamma B+C)}(W_\ell-P_\ell)
\nonumber\\[-2truemm]
&\qquad\qquad\qquad\quad
{}+\sum_\ell(W_\ell-P_\ell)W_\ell^{-1}\rme^{t(\gamma B+C)}W_\ell.
\end{align}
Note the intertwining relations
\begin{gather}
W_\ell=W_\ell P_\ell=\tilde{P}_\ell W_\ell,
\\
W_\ell^{-1}=P_\ell W_\ell^{-1}=W_\ell^{-1}\tilde{P}_\ell
\end{gather}
in~(\ref{eqn:IntertwiningW}) and~(\ref{eqn:IntertwiningWinv}).
Recall here the definitions of $W_\ell$ and $W_\ell^{-1}$ in~(\ref{eqn:SW}) and~(\ref{eqn:SWinv}), and the pseudoinverse $(\tilde{U}_\ell U_\ell)^{-1}$ in~(\ref{eqn:pseudoinv}).
Since 
\begin{gather}
U_\ell=U_\ell P_\ell,\qquad
P_\ell U_\ell=P_\ell,\\
\tilde{U}_\ell=P_\ell\tilde{U}_\ell,\qquad
\tilde{U}_\ell P_\ell=P_\ell,
\end{gather}
as noted in~(\ref{eqn:UPPU}) and~(\ref{eqn:UPPURev}), we have
\begin{gather}
W_\ell
=[1+(U_\ell-P_\ell)][1+(\tilde{U}_\ell-P_\ell)(U_\ell-P_\ell)]^{-1/2}P_\ell,
\displaybreak[0]\\
W_\ell^{-1}
=P_\ell[1+(\tilde{U}_\ell-P_\ell)(U_\ell-P_\ell)]^{-1/2}[1+(\tilde{U}_\ell-P_\ell)],
\end{gather}
and
\begin{align}
&W_\ell-P_\ell
\nonumber
\\
&\quad
=[1+(U_\ell-P_\ell)][1+(\tilde{U}_\ell-P_\ell)(U_\ell-P_\ell)]^{-1/2}-1,
\displaybreak[0]
\\
&W_\ell^{-1}-P_\ell
\nonumber\\
&\quad
=[1+(\tilde{U}_\ell-P_\ell)(U_\ell-P_\ell)]^{-1/2}[1+(\tilde{U}_\ell-P_\ell)]-1.
\end{align}
These are bounded by
\begin{gather}
\|W_\ell\|,\|W_\ell^{-1}\|
\le
\frac{1+\Theta_\ell}{\sqrt{1-\Theta_\ell^2}}\|P_\ell\|,
\\
\|W_\ell-P_\ell\|,\|W_\ell^{-1}-P_\ell\|
\le
\frac{1+\Theta_\ell}{\sqrt{1-\Theta_\ell^2}}-1,
\end{gather}
using the bounds $\|U_\ell-P_\ell\|\le\Theta_\ell$ and $\|\tilde{U}_\ell-P_\ell\|\le\Theta_\ell$ in~(\ref{eqn:Ubound}) and~(\ref{eqn:UboundRev}).
We hence get
\begin{align}
&
\bigl\|\rme^{t(\gamma B+C)}-\rme^{t(\gamma B+K)}\bigr\|
\nonumber
\\
&\quad
\le
\sum_\ell
\Bigl(
\|W_\ell-P_\ell\|
+\|(W_\ell-P_\ell)W_\ell^{-1}\|\|W_\ell\|
\Bigr)
\nonumber\\[-2truemm]
&\qquad\qquad\qquad\qquad\qquad\qquad\qquad\quad\ \ %
{}\times
\|\rme^{t(\gamma B+C)}\|
\nonumber
\displaybreak[0]
\\
&\quad
\le
\sum_\ell
\frac{2}{1-\Theta_\ell}
\left(
\sqrt{\frac{1+\Theta_\ell}{1-\Theta_\ell}}-1
\right)
\|P_\ell\|\|\rme^{t(\gamma B+C)}\|
\nonumber
\displaybreak[0]
\\
&\quad
=
\sum_\ell
\left(
\frac{1}{\sqrt{1-\gamma_\ell/\gamma}}+1
\right)
\left(
\frac{1}{\sqrt[4]{1-\gamma_\ell/\gamma}}-1
\right)
\|P_\ell\|
\nonumber\\
&\qquad\qquad\qquad\qquad\qquad\qquad\qquad\qquad\quad%
{}\times
\|\rme^{t(\gamma B+C)}\|,
\label{eqn:DistanceBoundK}
\end{align}
where 
\begin{equation}
\gamma_\ell = 4 \|S_\ell\|\|C\|\|P_\ell\| \frac{1-(\|S_\ell\|\|N_\ell\|)^{n_\ell}}{1-\|S_\ell\|\|N_\ell\|}.
\label{eq:gammaell}
\end{equation}
This can be loosely bounded as in~(\ref{eqn:EternalBoundK}) for $\gamma\ge2 \max_\ell \gamma_\ell$, in the norm induced by the operator trace norm.

The distance between $\rme^{t(\gamma B+C)}$ and $\rme^{t(\gamma B+D)}$, which are similar to each other through $U$, can be bounded in a similar way.
Note the intertwining relations
\begin{gather}
U_\ell=U_\ell P_\ell=\tilde{P}_\ell U_\ell,
\\
U_\ell^{-1}=P_\ell U_\ell^{-1}=U_\ell^{-1}\tilde{P}_\ell,
\end{gather}
where
\begin{equation}
U_\ell^{-1}
=(\tilde{U}_\ell U_\ell)^{-1}\tilde{U}_\ell
\end{equation}
is a pseudoinverse satisfying 
\begin{equation}
U_\ell^{-1}U_\ell=P_\ell,\qquad
U_\ell U_\ell^{-1}=\tilde{P}_\ell.
\end{equation}
It is bounded by
\begin{equation}
\|U_\ell^{-1}\|
\le\frac{1+\Theta_\ell}{1-\Theta^2_\ell}.
\end{equation}
Then, the difference
\begin{align}
\rme^{t(\gamma B+C)}
-
\rme^{t(\gamma B+D)}
={}&
{-\sum_\ell\rme^{t(\gamma B+C)}}
(U_\ell-P_\ell)
\nonumber\\
&
{}+\sum_\ell(U_\ell-P_\ell)U_\ell^{-1}\rme^{t(\gamma B+D)}U_\ell
\end{align}
is bounded by
\begin{align}
&
\bigl\|\rme^{t(\gamma B+C)}-\rme^{t(\gamma B+D)}\bigr\|
\nonumber
\displaybreak[0]\\
&\quad
\le
\sum_\ell
\Bigl(
\|U_\ell-P_\ell\|
+\|(U_\ell-P_\ell)U_\ell^{-1}\|\|U_\ell\|
\Bigr)
\nonumber\\[-2truemm]
&\qquad\qquad\qquad\qquad\qquad\qquad\qquad\quad\ \ %
{}\times
\|\rme^{t(\gamma B+C)}\|
\nonumber\\
&\quad
\le
\sum_\ell
\frac{2\Theta_\ell}{1-\Theta_\ell}
\|P_\ell\|\|\rme^{t(\gamma B+C)}\|
\nonumber\\
&\quad
=
\sum_\ell
\left(
\frac{1}{\sqrt{1-\gamma_\ell/\gamma}}-1
\right)
\|P_\ell\|
\|\rme^{t(\gamma B+C)}\|.
\end{align}
This bound is smaller than the bound on the distance $\|\rme^{t(\gamma B+C)}-\rme^{t(\gamma B+K)}\|$ in~(\ref{eqn:DistanceBoundK}).

Since $1/\sqrt{1-x} -1 < x$ for $0<x \leq 1/2$, this can be loosely bounded as in~(\ref{eqn:EternalBoundD}) for $\gamma\ge2 \max_\ell \gamma_\ell$, in the 1-1 norm induced by the operator trace norm, 
\begin{equation}
\| B\| = \sup_{\|\varrho\|_1 =1} \|B(\varrho)\|_1,
\end{equation}
where $\|\varrho\|_1 = \tr |\varrho |$.

Moreover, in the unitary case, by using the spectral norm, so that $\|A\|=\|A^\dag A\|^{1/2}=\|AA^\dag\|^{1/2}$, tighter bounds are available.
For instance, by using the unitarity of $W$ and $\rme^{t(\gamma B+C)}$, whose  norms are $\|W\|=\|\rme^{t(\gamma B+C)}\|=1$, and the orthogonality $(W_k-P_k)(W_\ell-P_\ell)^\dag=0$ for $k\neq\ell$, we can bound the distance as
\begin{align}
&\bigl\|
\rme^{t(\gamma B+C)}
-
\rme^{t(\gamma B+K)}
\bigr\|
\nonumber
\\
&\quad
=
\bigl\|
{-\rme^{t(\gamma B+C)}(W-I)}
+(W-I)W^{-1}\rme^{t(\gamma B+C)}W
\bigr\|
\nonumber
\displaybreak[0]
\\
&\quad
\le
2\|W-I\|
\vphantom{\biggr)}
\nonumber
\displaybreak[0]
\\
&\quad
=
2\biggl\|
\sum_\ell(W_\ell-P_\ell)
\biggr\|
\nonumber
\displaybreak[0]
\\
&\quad
=
2\biggl\|
\sum_k(W_k-P_k)
\sum_\ell(W_\ell-P_\ell)^\dag
\biggr\|^{1/2}
\nonumber
\displaybreak[0]
\\
&\quad
=
2\biggl\|
\sum_\ell(W_\ell-P_\ell)(W_\ell-P_\ell)^\dag
\biggr\|^{1/2}
\nonumber\\
&\quad
\le
2\biggl(
\sum_\ell\|W_\ell-P_\ell\|^2
\biggr)^{1/2}
\nonumber
\\
&\quad
\le
2\sqrt{
\sum_\ell\biggl(
\sqrt{\frac{1+\Theta_\ell}{1-\Theta_\ell}}-1
\biggr)^2
}
\nonumber
\displaybreak[0]
\\
&\quad
\le
2\sqrt{d}
\max_\ell\biggl(
\sqrt{\frac{1+\Theta_\ell}{1-\Theta_\ell}}-1
\biggr)
\nonumber
\displaybreak[0]
\\
&\quad
=
2\sqrt{d}\,
\biggl(
\frac{1}{\sqrt[4]{1-4\|C\|/(\gamma\eta)}}-1
\biggr),
\label{eqn:DistanceBoundKunitary}
\end{align}
where $d$ is the number of distinct eigenvalues of $B$, and 
\begin{equation}
\eta=\min_{k\neq\ell}|b_k-b_\ell|
\end{equation}
is the spectral gap of $B$.
Note that $\mu_\ell=1$, $\|P_\ell\|=1$, and hence $\gamma_\ell=4\|S_\ell\|\|C\| \le4\|C\|/ \eta$ in the unitary case.

For the distance between $\rme^{t(\gamma B+C)}$ and $\rme^{t(\gamma B+D)}$, the similarity transformation $U$ between them is not unitary even for unitary evolution, but anyway, we can bound it as
\begin{align}
&\bigl\|
\rme^{t(\gamma B+C)}
-
\rme^{t(\gamma B+D)}
\bigr\|
\nonumber\\
&\ 
=
\bigl\|
{-\rme^{t(\gamma B+C)}(U-I)}
+(U-I)U^{-1}\rme^{t(\gamma B+C)}U
\bigr\|
\nonumber
\displaybreak[0]
\\
&\ 
\le
\|U-I\|
+
\bigl\|(U-I)U^{-1}\rme^{t(\gamma B+C)}U\bigr\|
\vphantom{\biggr)}
\nonumber
\displaybreak[0]
\\
&\ 
=
\biggl\|
\sum_\ell(U_\ell-P_\ell)
\biggr\|
+
\biggl\|
\sum_\ell(U_\ell-P_\ell)U_\ell^{-1}\rme^{t(\gamma B+C)}U_\ell
\biggr\|
\nonumber
\displaybreak[0]
\\
&\ 
\le
\biggl(
\sum_\ell\|U_\ell-P_\ell\|^2
\biggr)^{1/2}
\nonumber
\displaybreak[0]
\\
&\qquad\qquad\qquad
{}+
\biggl(
\sum_\ell\bigl\|(U_\ell-P_\ell)U_\ell^{-1}\rme^{t(\gamma B+C)}U_\ell\bigr\|^2
\biggr)^{1/2}
\nonumber
\displaybreak[0]
\\
&\ 
\le
\sqrt{
\sum_\ell\Theta_\ell^2
}
+
\sqrt{
\sum_\ell\biggl(
\Theta_\ell\frac{1+\Theta_\ell}{1-\Theta_\ell}
\biggr)^2
}
\nonumber
\displaybreak[0]
\\
&\ 
\le
\sqrt{d}
\max_\ell\left(
\frac{2\Theta_\ell}{1-\Theta_\ell}
\right)
\nonumber
\displaybreak[0]
\\
&\ 
=
\sqrt{d}\,
\biggl(
\frac{1}{\sqrt{1-4\|C\|/(\gamma\eta)}}-1
\biggr),
\end{align}
where we have used the orthogonality $U_k U_\ell^\dag=0$ for $k\neq\ell$.
This bound is larger than the bound on the distance $\|\rme^{t(\gamma B+C)}-\rme^{t(\gamma B+K)}\|$ in~(\ref{eqn:DistanceBoundKunitary}).


\begin{thebibliography}{10}
\bibitem{note:GKLS}
Strictly speaking, the following perturbative terms do not fit in the GKLS
  form, since some of their Kossakowski matrices are not positive-semidefinite,
  and should not be called GKLS form. We will however call the
  form~(\ref{eqn:GKLS}) GKLS form.
\end{thebibliography}
\end{document}